\documentclass[twocolumn]{aastex6}
\usepackage{amssymb}
\usepackage{xcolor}
\usepackage{amsmath}
\begin{document}


\title{Deflating Super-Puffs: Impact of Photochemical Hazes on the Observed Mass-Radius Relationship of Low Mass Planets}


\author{Peter Gao\altaffilmark{1,2} and Xi Zhang\altaffilmark{3}}

\altaffiltext{1}{Department of Astronomy, University of California, Berkeley, CA 94720; \href{mailto:gaopeter@berkeley.edu}{gaopeter@berkeley.edu}}
\altaffiltext{2}{51 Pegasi b Fellow}
\altaffiltext{3}{Department of Earth and Planetary Sciences, University of California, Santa Cruz, CA 95064}




\altaffiltext{1}{51 Pegasi b Fellow}
\altaffiltext{2}{gaopeter@berkeley.edu}

\begin{abstract}
The observed mass-radius relationship of low-mass planets informs our understanding of their composition and evolution. Recent discoveries of low mass, large radii objects (``super-puffs'') have challenged theories of planet formation and atmospheric loss, as their high inferred gas masses make them vulnerable to runaway accretion and hydrodynamic escape. Here we propose that high altitude photochemical hazes could enhance the observed radii of low-mass planets and explain the nature of super-puffs. We construct model atmospheres in radiative-convective equilibrium and compute rates of atmospheric escape and haze distributions, taking into account haze coagulation, sedimentation, diffusion, and advection by an outflow wind. We develop mass-radius diagrams that include atmospheric lifetimes and haze opacity, which is enhanced by the outflow, such that young ($\sim$0.1-1 Gyr), warm (T$_{eq}$ $\geq$ 500 K), low mass objects ($M_c$ $<$ 4M$_{\Earth}$) should experience the most apparent radius enhancement due to hazes, reaching factors of three. This reconciles the densities and ages of the most extreme super-puffs. For Kepler-51b, the inclusion of hazes reduces its inferred gas mass fraction to $<$10\%, similar to that of planets on the large radius side of the sub-Neptune radius gap. This suggests that Kepler-51b may be evolving towards that population, and that some warm sub-Neptunes may have evolved from super-puffs. Hazes also render transmission spectra of super-puffs and sub-Neptunes featureless,  consistent with recent measurements. Our hypothesis can be tested by future observations of super-puffs' transmission spectra at mid-infrared wavelengths, where we predict that the planet radius will be half of that observed in the near-infrared. 
\end{abstract}




\section{Introduction}\label{sec:intro}

The \textit{Kepler} mission revealed that planets with radii between that of Earth and Neptune are the most abundant in the Galaxy \citep{borucki2011,howard2012,dressing2013,fressin2013,petigura2013}. These worlds, which we will refer to collectively as ``sub-Neptunes'', are divided into two populations in radii by a ``valley'' centered at $\sim$1.8R$_{\Earth}$, where there is a dearth of planets \citep{fulton2017,vaneylen2018}. Combining precise radii measured by \textit{Kepler} with mass measurements from radial velocities and transit-timing variations, the smaller radii population ($\leq$1.5R$_{\Earth}$) has been found to possess composition similar to that of Earth \citep{dressing2015}, while the larger radii population (2R$_{\Earth}\leq R \leq$3R$_{\Earth}$) is likely composed of rocky cores surrounded by a gas envelope with a mass that is a few $\%$ of the core \citep{owen2017}. This has been interpreted as a signature of atmospheric loss through photoevaporation due to extreme ultraviolet irradiation from these planets' host stars \citep{lopez2013,owen2013,jin2014,chen2016,owen2016}, and/or high internal luminosity stemming from cooling of the rocky core \citep{ginzburg2018}. An alternate hypothesis states that at least some of the larger radii population are water worlds \citep[e.g.][]{zeng2019}, where a core of rock and ice underlies a thin gas layer.

Another revelation produced by \textit{Kepler} is the existence of ``super-puffs,'' temperate ($T_{eq}$ $\sim$300-800 K) worlds that are larger than the large radii population of sub-Neptunes ($\geq$4R$_{\Earth}$), but with similar masses ($\leq$10M$_{\Earth}$). The resulting low densities lead to inferred gas mass fractions $>$10$\%$ \citep{lopez2014}, significantly greater than that inferred for the large radii population of sub-Neptunes. The relative rarity of super-puffs in our current sample of discoveries \citep[see Table \ref{table:superpuffs} and e.g.][]{masuda2014,jontofhutter2014,ofir2014,mills2016} begs the question: Are they a separate population of planets with unique formation and evolutionary histories, or are they related to the much more numerous sub-Neptunes? Indeed, their large apparent gas mass fractions present a puzzle for planet formation theories \citep{ikoma2012,inamdar2015}. \citet{lee2016} posit that, in order for super-puffs to exist at their current close-in orbits among sub-Neptunes (P $<$ 100 days), they must have formed beyond 1 AU with dust-free atmospheres and migrated inwards. This is consistent with super-puffs being on the outer parts of resonant chains, which is evidence of migration \citep{lee2016}, though this is not always the case, such as for Kepler-79d \citep{jontofhutter2014}, which is sandwiched between more typical sub-Neptunes. Meanwhile, \citet{millholland2019} showed that the large radii of some super-puffs could be sustained by tidal heating through obliquity tides, which increases the internal entropy of these planets. For some super-puffs, however, their large inferred gas mass fractions also present a problem for their continued existence. \citet{wang2019} showed that, if transit observations probed purely gaseous, clear atmospheres, then Kepler-51b, one of the least dense super-puffs (Table \ref{table:superpuffs}), would possess an atmospheric loss timescale on the order of only 10$^3$ years due to hydrodynamic boil-off; this is a significantly shorter timescale than the inferred age of the system \citep[0.3 Gyr;][]{masuda2014}. 

\begin{deluxetable*}{lccccc}
\tablecolumns{6}
\tablecaption{Properties of observed super-puffs\tablenotemark{*}\label{table:superpuffs} }
\tablehead{
\colhead{Planet} & \colhead{Mass (M$_{\Earth}$)} & \colhead{Radius (R$_{\Earth}$)\tablenotemark{+}} & \colhead{Density (g cm$^{-3}$)} & \colhead{$T_{eq}$ (K)} & \colhead{Age (Gyr)}}
\startdata
Kepler-11 e  &  8.0 $^{+ 1.5 }_{ -2.1 }$  &  4.19 $^{+ 0.07 }_{ -0.09 }$  &  0.6 $^{+ 0.1 }_{ 0.2 }$  &  641  &  8.5 $^{+ 1.1 }_{ -1.4 }$ \\
Kepler-177 c  &  7.5 $^{+ 3.5 }_{ -3.1 }$  &  7.1 $^{+ 3.71 }_{ -0.72 }$  &  0.1 $^{+ 0.2 }_{ 0.06 }$  &  511  &  4.37$^{+ 3.63 }_{ -2.55 }$ \\
Kepler-223 d  &  8.0 $^{+ 1.5 }_{ -1.3 }$  &  5.24 $^{+ 0.26 }_{ -0.45 }$  &  0.31 $^{+ 0.07 }_{ 0.09 }$  &  791  &  4.27$^{+ 3.92 }_{ -2.48 }$ \\
Kepler-223 e  &  4.8 $^{+ 1.4 }_{ -1.2 }$  &  4.6 $^{+ 0.27 }_{ -0.41 }$  &  0.27 $^{+ 0.09 }_{ 0.1 }$  &  719  &  4.27$^{+ 3.92 }_{ -2.48 }$ \\
Kepler-47 c  &  3.17 $^{+ 2.18 }_{ -1.25 }$  &  4.65 $^{+ 0.09 }_{ -0.07 }$  &  0.17 $^{+ 0.12 }_{ 0.07 }$  &  260  &  1.65$^{+ 0.02 }_{ -0.01 }$ \\
Kepler-51 b  &  3.69 $^{+ 1.86 }_{ -1.59 }$  &  6.89 $^{+ 0.14 }_{ -0.14 }$  &  0.064 $^{+ 0.024 }_{ 0.024 }$  &  500  &  0.3 $^{+ 2.3 }_{ -2.3 }$ \\
Kepler-51 c  &  4.43 $^{+ 0.54 }_{ -0.54 }$  &  8.98 $^{+ 2.84 }_{ -2.84 }$  &  0.034 $^{+ 0.069 }_{ 0.019 }$  &  404  &  0.3 $^{+ 2.3 }_{ -2.3 }$ \\
Kepler-51 d  &  5.7 $^{+ 1.12 }_{ -1.12 }$  &  9.46 $^{+ 0.16 }_{ -0.16 }$  &  0.038 $^{+ 0.006 }_{ 0.006 }$  &  351  &  0.3 $^{+ 2.3 }_{ -2.3 }$ \\
Kepler-79 d  &  6.0 $^{+ 2.1 }_{ -1.6 }$  &  7.16 $^{+ 0.13 }_{ -0.16 }$  &  0.09 $^{+ 0.03 }_{ 0.02 }$  &  640  &  3.44 $^{+ 0.6 }_{ -0.91 }$ \\
Kepler-87 c  &  6.4 $^{+ 0.8 }_{ -0.8 }$  &  6.14 $^{+ 0.29 }_{ -0.29 }$  &  0.15 $^{+ 0.03 }_{ 0.03 }$  &  394  &  7.5 $^{+ 0.5 }_{ -0.5 }$ \\
\enddata
\tablenotetext{*}{Masses, radii, and ages obtained from NASA Exoplanet Archive. Densities calculated from masses and radii. $T_{eq}$ computed from luminosities and semimajor axes obtained from NASA Exoplanet Archive. Mass, radius, and densities for the Kepler-51 planets obtained from \citet{libbyroberts2019}.}
\tablenotetext{+}{In the \textit{Kepler} Bandpass.}
\end{deluxetable*}

The longevity and the large inferred gas mass fractions of super-puffs may be reconciled if their \textit{Kepler}-derived radii are probing significantly lower pressures than previously assumed. In the \textit{Kepler} bandpass ($\sim$430-880 nm), a clear, H$_2$/He-dominated atmosphere becomes opaque in transmission at $\sim$100 mbar \citep{hubbard2001,lammer2016}, though this will vary depending on planet equilibrium temperature, gravity, and atmospheric metallicity. However, if unknown opacity sources in addition to gaseous absorbers were present, then the pressures probed--and the inferred gas mass fraction--could be significantly lower. Since the atmospheric loss rate is a sensitive function of the atmospheric density at the exobase, which is related to the atmosphere mass, knowing the pressures probed in transmission is vital for computing atmospheric lifetimes. 

An important but uncertain opacity source in planetary atmospheres is aerosols, which are widespread in exoplanet atmospheres across planet temperatures, masses, and ages \citep{sing2016,crossfield2017}. Recent \textit{Hubble Space Telescope} observations of the transmission spectra of Kepler-51b and d using the G141 grism on Wide Field Camera 3 ($\sim$1.1-1.7 $\mu$m) showed them to be flat \citep{libbyroberts2019}, despite the presence of a strong water band at those wavelengths, suggesting the presence of high altitude aerosols. Importantly, the lack of detection of any molecular features means that only an upper limit can be placed on the pressures probed in transmission, and thus the clear atmosphere radius of the planet. As most planets observed by \textit{Kepler} lack radius measurements at other wavelengths, high altitude hazes cannot be ruled out, thus calling into question their clear atmosphere radii. Furthermore, this same issue can plague discoveries by the \textit{Transiting Exoplanet Survey Satellite} (TESS), motivating us to evaluate the impact of high altitude hazes on the observed radius of sub-Neptunes and super-puffs. 

It is nontrivial to sustain aerosols at the low pressures ($\sim$1 $\mu$bar) needed to create flat transmission spectra and significantly alter a planet's observed radius. Aerosols that form through condensation of atmospheric trace gases--clouds--are difficult to loft to low pressures, as they are fueled by upwelling of condensate vapor near the cloud base deeper in the atmosphere \citep{ackerman2001,powell2018,gao2018a}. \citet{gao2018b} used an aerosol microphysics model to generate KCl clouds in the atmosphere of the sub-Neptune GJ 1214b in an attempt to explain its flat transmission spectrum \citep{kreidberg2014}, and were only successful when the atmospheric metallicity was high (1000$\times$solar) and the strength of vertical mixing was several orders of magnitude greater than predicted by general circulation models \citep{charnay2015a}. Using the predicted strength of vertical mixing, \citet{ohno2018} was unable to match the inferred cloud top pressure with KCl clouds even at high metallicities, unless the cloud particles were porous aggregates \citep{ohno2019}. 

In comparison, aerosols that form at low pressures through the actions of photochemistry--hazes--have an advantage since there is no need for lofting. For example, optical transits of Saturn's moon Titan probe altitudes upwards of 300 km (10-100 $\mu$bar) above the surface due to opacity from photochemical hazes \citep{robinson2014}, while haze formation occurs at pressures as low as 0.1 nbar at an altitude of 1000 km, which is comparable to the solid body radius of Titan of 2575 km \citep{horst2017}. \citet{morley2013,morley2015} showed that a flat spectrum for GJ 1214b can be generated using photochemical hazes if $\geq$10\% of the products of methane and nitrogen photolysis are converted into hazes with particle radius $\sim$0.1 $\mu$m. \citet{adams2019} were able to reproduce GJ 1214b's spectrum using a haze microphysics model, a parameterized haze production rate, the predicted strength of vertical mixing from general circulation models, and fractal aggregate haze particles; they also showed that hazy transmission spectra are able to probe low pressures near 1 $\mu$bar given sufficiently high haze production rates. A number of recent works have combined both detailed photochemical simulations and haze microphysics \citep{kawashima2018,kawashima2019a,kawashima2019b,lavvas2019}, and showed that hazes on low mass planets including Kepler-51b could block transmission of stellar photons for pressures greater than 0.1-1 $\mu$bar. In particular, \citet{kawashima2019a} showed that the radii of Kepler-51b observed by \textit{Kepler} can be reproduced by a model atmosphere with a radius at 1000 bars of only 1.8R$_{\Earth}$ and high altitude hazes, suggesting that the gas mass fractions of super-puffs could be drastically lower than previously inferred. 

Few previous works investigated the impact of high altitude hazes on the observed radii of low mass planets. \citet{lammer2016} suggested that high altitude aerosol particles of unknown composition and origin hid the ``true'' radius of the hot Neptune CoRoT-24b, which, like Kepler-51b, would have lost its atmosphere long ago given its observed mass, radius, and stellar irradiation levels. However, they did not quantify the processes by which the aerosols may be sustained at the altitudes needed to explain the observations. \citet{wang2019} argued that 10 ${\rm \AA}$ dust grains--such as tiny graphite particles and polycyclic aromatic hydrocarbons--entrained in hydrodynamic outflows can be carried to low pressures, thereby explaining the large radii of super-puffs. However, they did not quantitatively evaluate the growth of the particles via microphysical processes during transport, which could impede lofting (Ohno \& Tanaka, in prep). If growth were inhibited by some unknown process and the particles remained small in the outflow, as proposed by \citet{wang2019}, then it would be essential to evaluate whether grains as small as 10 \AA~can reproduce the flat near-IR transmission spectrum of Kepler-51b observed by  \citet{libbyroberts2019}.

In this work, we combine the physics of atmospheric escape and aerosol evolution to explore the impact of photochemical hazes on the observed radii of sub-Neptunes and super-puffs. We construct a grid of model atmospheres in radiative-convective equilibrium for various core and atmosphere masses, intrinsic luminosities, and equilibrium temperatures, and compute their atmospheric lifetimes and outflow rates. We then use a one-dimensional (1D) aerosol microphysics model to characterize the vertical and size distribution of photochemical hazes in these atmospheres under the influence of the outflows. We evaluate the effects such hazes have on the observed radius of these planets by calculating the transmission spectrum of their atmospheres with and without hazes. 

Our efforts allow us to produce ``hazy'' mass-radius diagrams of low-mass planets that take into account atmospheric lifetimes. Mass-radius diagrams have long been used to estimate the composition of planets \citep{stevenson1982}, with renewed interest in the era of exoplanets \citep{seager2007,fortney2007,rogers2011,mordasini2012,zeng2016,zeng2019}. Comparisons of the observed masses and radii of the growing population of worlds beyond our Solar System with standard mass-radius diagrams have shown possible transitions in composition with increasing mass from pure rocks to rocky planets with gas envelopes (or water worlds) to Neptune-like worlds to gas giants \citep{rogers2015,chen2017}. Here, our model framework allows us to compute the radius of the planet as a function of wavelength, as controlled by gas and aerosol opacity in the atmosphere, and link each mass and radius to the lifetime of the associated atmosphere due to escape. This adds two additional dimensions to the standard mass-radius diagram: the atmospheric composition, including aerosols, and the evolutionary timescale of the planets.


In ${\S}$\ref{sec:theory}, we detail the construction of our grid of radiative-convective atmospheric models and the computation of their lifetimes, along with our treatment of haze microphysics. We present our results in ${\S}$\ref{sec:results}, where we show how hazes evolve with increasing outflow speeds, and how such an evolution impacts the radius enhancement due to hazes and the flattening of near-infrared transmission spectra. We also present our hazy mass-radius diagrams. We discuss the implications of our work in ${\S}$\ref{sec:discussion}, including how certain super-puffs and sub-Neptunes are related, the transmission spectra of hazy worlds in the mid-infrared, and the importance of haze opacity to the radiative transfer of the atmosphere. We also investigate the sensitivity of our results to assumptions of eddy diffusivity, metallicity, and haze production efficiency. We summarize our findings and present our conclusions in ${\S}$\ref{sec:conclusions}. 

\section{Theory}\label{sec:theory}


\subsection{Atmospheric Structure}\label{sec:radcon}

We begin by constructing a grid of model sub-Neptunes by assuming a simplified structure of an atmosphere overlying a rocky core. We vary the core mass $M_c$, atmosphere mass $M_a$, equilibrium temperature $T_{eq}$, and intrinsic temperature $T_{int}$, with which we parameterize the internal luminosity of these planets, $L_p$, through

\begin{equation}
\label{eq:tint}
L_p = 4\pi r_{rcb}^2 \sigma T_{int}^4,
\end{equation}

\noindent where $\sigma$ is the Stefan-Boltzmann constant and $r_{rcb}$ is the radius of the radiative convective boundary (RCB), assumed to coincide with the photosphere of the planet for simplicity. We choose $T_{eq}$ = 700, 500, and 300 K to emulate the equilibrium temperatures of observed super-puffs (Table \ref{table:superpuffs}), and $T_{int}$ = 75 K and 30 K to cover variations in age. Given the evolutionary curves of \citet{lopez2014}, we use $T_{int}$ = 75 K to represent $\sim$0.1-3 Gyr old planets, and $T_{int}$ = 30 K to represent $>$3 Gyr old planets, though we note that $T_{int}$ is higher for higher mass planets than lower mass planets of the same age. The core mass is varied from 1.5M$_{\oplus}$ to 7.5M$_{\oplus}$, and the core radius $r_c$ is computed assuming an Earth-like (32.5\% Fe+67.5\% MgSiO$_3$) mass-radius relationship taken from \citet{zeng2019}. The atmospheric mass is varied from 0.1\% to 30\% of the core mass, capturing the inferred atmospheric mass fractions of sub-Neptunes and super-puffs \citep{lopez2014}. 

For each model planet, we construct temperature-pressure profiles of the atmosphere  \citep[e.g.][]{rafikov2006,piso2014,owen2016,wang2018}. The model atmospheres extend from the rocky core to 1 nbar to capture the haze formation region. We will use these atmospheres to simulate haze distributions and transmission spectra. 

We first solve for the convective region by assuming hydrostatic equilibrium,

\begin{equation}
\label{eq:hydro}
\frac{dP}{dr} = -\rho\frac{GM_c}{r^2},
\end{equation}

\noindent where $P$ is pressure, $r$ is distance from the center of the planet, $\rho$ is atmospheric mass density, and $G$ is the gravitational constant. By only using $M_c$ in defining the gravity, we ignore the self-gravity of the atmosphere, which is a valid assumption as long as the atmospheric mass is less than a few tens of \% of the core mass \citep{rafikov2006,piso2014,owen2017}. We assume the ideal gas equation of state, using a polytrope,

\begin{equation}
\label{eq:poly}
P = \rho \frac{R}{\mu} T = K\rho^{\gamma},
\end{equation}

\noindent where $R$ is the gas constant in J mol$^{-1}$ K$^{-1}$, $\mu$ is the mean molecular weight of the atmosphere, $T$ is temperature, $K$ is a constant, and $\gamma$ is the adiabatic index of the gas, assumed to be $7/5$ for a largely molecular H$_2$ atmosphere. The density structure $\rho(r)$ of the convective region can be computed by combining Eqs. \ref{eq:hydro} and \ref{eq:poly}, yielding \citep[e.g.][]{owen2017},

\begin{equation}
\label{eq:convrho}
\rho(r) = \rho_{rcb} \left [ 1 + \nabla_{ad}\frac{\mu GM_c}{RT_{eq}} \left (\frac{1}{r} - \frac{1}{r_{rcb}} \right ) \right ]^{\frac{1}{\gamma - 1}},
\end{equation}

\noindent where $\rho_{rcb}$ is the atmospheric mass density at the RCB, and we have set the temperature there to $T_{eq}$. $\nabla_{ad}$ is the adiabatic temperature gradient given by,

\begin{equation}
\label{eq:adgrad}
\nabla_{ad} = \left (\frac{d{\rm Ln}T}{d{\rm Ln}P} \right )_{ad} = \frac{\gamma - 1 }{\gamma}.
\end{equation}

\noindent The atmospheric mass can be obtained by integrating over the convective region from $r_c$ to $r_{rcb}$, where most of the atmospheric mass is concentrated,  

\begin{equation}
\label{eq:mass}
M_a = \int_{r_c}^{r_{rcb}} 4\pi r^2 \rho(r) dr,
\end{equation}

\noindent and the temperature gradient at $r_{rcb}$ must transition smoothly from an adiabatic gradient to a radiative gradient, given by \citet{piso2014} as

\begin{equation}
\label{eq:rdgrad}
\nabla_{rd} = \frac{3\kappa P L_p}{64 \pi G M_c \sigma T^4},
\end{equation}

\noindent where $L_p$ is given by Eq. \ref{eq:tint} and we have replaced the mass enclosed within a radius corresponding to pressure $P$ with $M_c$. $\kappa$ is the local opacity of the atmosphere, for which we use the opacity tables of \citet{freedman2014}, assuming a solar metallicity atmosphere ($\mu$ = 2.3559 g mol$^{-1}$). While the metallicity of the atmosphere may be higher (see ${\S}$\ref{sec:metallicity}), current observations do not provide sufficient constraints on the atmospheric metallicity of exo-Neptunes and sub-Neptunes \citep[e.g.][]{fraine2014,wakeford2017b,benneke2019,chachan2019}. 

\begin{figure*}[hbt!]
\centering
\includegraphics[width=0.8 \textwidth]{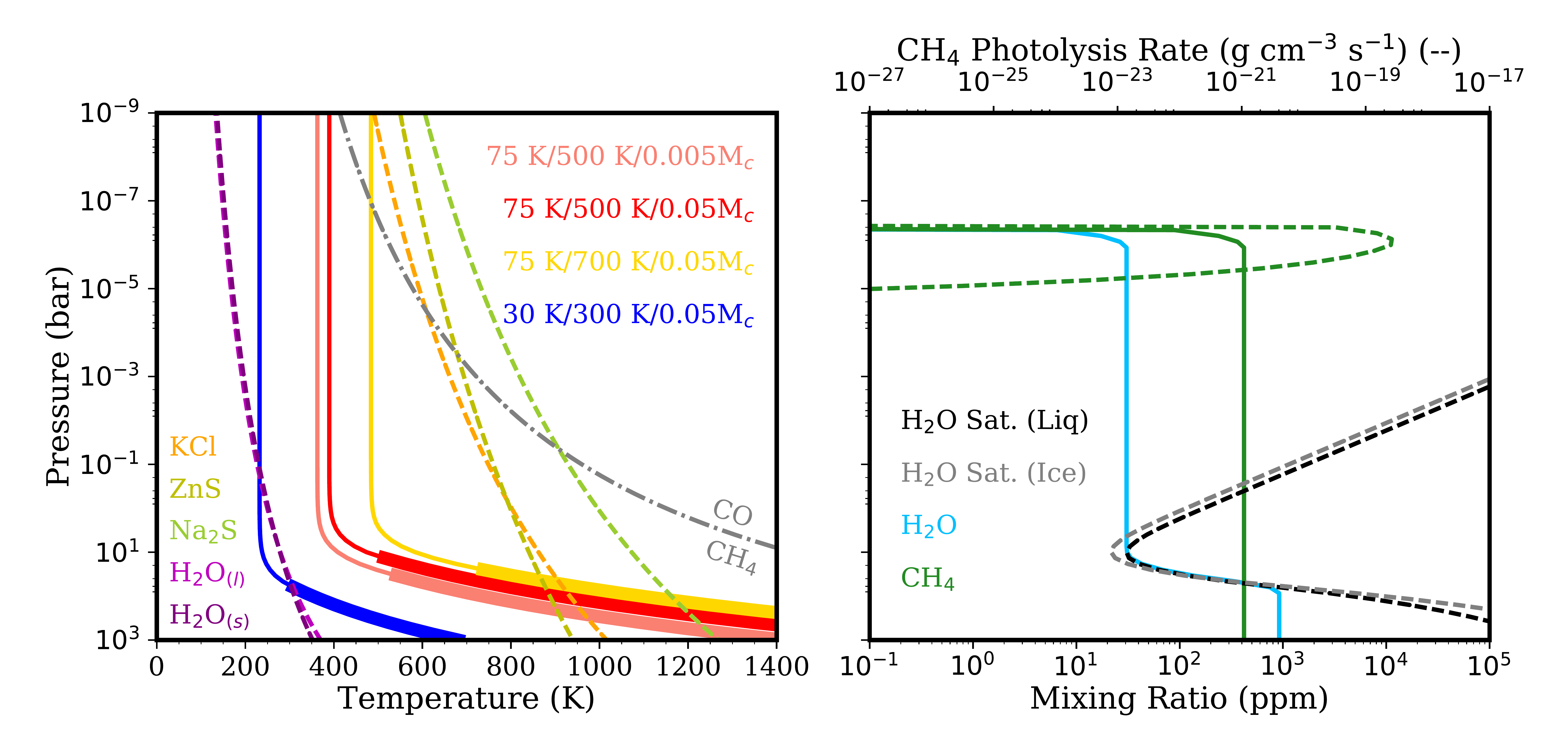}
\caption{(Left) Select temperature--pressure profiles for the given $T_{int}$/$T_{eq}$/$M_a$ cases, a core mass of 3M$_{\Earth}$, and solar metallicity, compared to the condensation curves of liquid water (magenta) and water ice (purple) \citep{murphy2005}, as well as KCl (orange), ZnS (yellow), and Na$_2$S (light green) \citep{morley2012}, and the transition curve for CO and CH$_4$ (gray) \citep{visscher2012}. The thicker curves mark the convective regions of the atmosphere. (Right) Profiles of water vapor (blue) and methane (green) mixing ratio for the $T_{eq}$ = 300 K, $T_{int}$ = 30 K, $M_c$ = 3M$_{\Earth}$, and $M_a$ = 0.05$M_c$ case, compared to the liquid water (black) and water ice (gray) saturation vapor mixing ratio}. The computed methane photolysis rate profile is shown in the green dashed curve (see ${\S}$\ref{sec:hazeprod}).
\label{fig:tprof}
\end{figure*}

The resulting temperature-pressure profiles all have the same general features: A deep adiabat in the convective region, a transition region around the RCB, and a radiative region that is nearly isothermal due to low $\kappa$ at pressures significantly lower than that at the RCB (Figure \ref{fig:tprof}). More massive atmospheres tend to be hotter, since the RCB is reached at lower gravity compared to that of lower mass atmospheres. From hydrostatic equilibrium, a lower gravity leads to a greater mass of atmosphere above any particular pressure, leading to higher opacity and thus higher temperatures. We neglect the effect of heating by high energy photons on the upper atmospheric temperature structure \citep{yelle2004}, as the magnitude and location of heating is uncertain, the objects in consideration receive lower stellar flux than highly irradiated hot Jupiters, and heating at such low pressures has little effect on the transmission spectra. The higher temperatures may affect the types of hazes that can form \citep{lavvas2017}, but we do not consider these effects in this study. 

Our method for constructing model atmospheres is successful for only part of the parameter space we explore. For low mass cores, high $T_{int}$ and $T_{eq}$, and high mass atmospheres, the RCB could be far enough away from the core (i.e. at low enough gravity) such that the gas opacity above the RCB is always sufficiently high to prevent the temperature profile from becoming radiative. Such scenarios are likely the result of our neglect of the atmosphere's self gravity, which would compress the atmosphere such that the local gravity would be higher. In addition, such models always possess atmospheric masses higher than models with extremely short lifetimes ($<$1 Myr; see ${\S}$\ref{sec:atmlife}), and so are unlikely to survive as sub-Neptunes for any significant amount of time. 

\subsection{Atmospheric Lifetime}\label{sec:atmlife}

Knowing the lifetimes of our model atmospheres is vital for assessing their relevance to the observed mass-radius diagram. We define the lifetime, $\tau_a$, as the total atmospheric mass divided by the atmospheric loss rate, the form of which is dependent on the loss regime. An important loss process is photoevaporation caused by X-rays and extreme ultraviolet (XUV) photons from the host star \citep[e.g.][]{owen2013,lopez2014}. The loss rate $\dot{M_e}$ due to photoevaporation in the energy-limited regime is, 

\begin{equation}
\label{eq:mdote}
\dot{M_e} = \epsilon \frac{\pi F_{XUV} R_w^3}{GM_c},
\end{equation}

\noindent where $\epsilon$ is an efficiency factor $\sim$10\% \citep{jackson2010,valencia2010,lopez2012,jin2014,chen2016,lopez2017} and $R_w$ is the planet radius at which the photoevaporative wind is launched, typically set to the XUV photosphere at $\sim$1 nbar \citep{murrayclay2009,lopez2017,wang2018}. $F_{XUV}$ is the flux of XUV radiation impacting the planet, which we take from Table 4 of \citet{ribas2005}; specifically, we sum the flux within the XUV wavelength range of 1-920 \AA$\:$ for stars closest in age to 0.1 Gyr ($T_{int}$ = 75 K) and 3 Gyr ($T_{int}$ = 30 K), yielding $\sim$121 (0.3 Gyr) and $\sim$14 (1.6 Gyr) ergs cm$^{-2}$ s$^{-1}$, respectively, which we then scale to the semi-major axes of the three $T_{eq}$ cases.

While energy-limited escape is applicable to every model atmosphere in our grid, it may not suffice for the lowest density worlds. \citet{owen2016} hypothesized that nascent planets with atmospheres extending to the bondi radius would experience a ``boil-off'' phase upon initial exposure to their host stars after shedding their protoplanetary disk shielding. Here, the energy driving escape is not XUV, but the bolometric luminosity of the star, $L_*$, and the binding energy released from gravitational contraction during boil-off, which keep the radiative region of the atmosphere isothermal. Following \citet{wang2018}, the bolometric luminosity component, $\dot{M_r}$, is 

\begin{equation}
\label{eq:mdotr}
\dot{M_r} = \frac{L_*}{4 \pi a^2}\pi r_{rcb}^2 \left (  \frac{2}{c_s^2} \right )
\end{equation}

\noindent where $c_s^2$ = $kT/m$ is the square of the isothermal sound speed. For sufficiently high energy input, the atmospheric loss rate is limited by $c_s$, resulting in an isothermal Parker wind mass loss rate \citep{parker1958}, $\dot{M_p}$, given by 

\begin{equation}
\label{eq:mdotp}
\dot{M_p} = 4 \pi r_s^2 c_s \rho_{rcb} \exp \left (\frac{3}{2}-\frac{2r_s}{r_{rcb}} \right ),
\end{equation}

\noindent where 

\begin{equation}
\label{eq:sonic}
r_s = \frac{GM_c}{2c_s^2},
\end{equation}

\noindent is the sonic radius. The product of $\rho_{rcb}$ and the exponential term in Eq. \ref{eq:mdotp} gives the density at the sonic radius, though we have neglected the outflow wind velocity term in the exponential, $-0.5v^2(r)/c_s^2$, where $v(r)$ is the outflow wind velocity at radius $r$ \citep{oklopcic2018}; for a sufficiently large $v(r)$, the hydrostatic assumption (Eq. \ref{eq:hydro}) breaks down, but this is not the case for most of our model grid. The mass loss rate of the boil-off, $\dot{M_b}$, is then 

\begin{equation}
\label{eq:boiloff}
\dot{M_b} = min\{\dot{M_r},\dot{M_p}\}.
\end{equation}

\begin{figure*}[hbt!]
\centering
\includegraphics[width=1.0 \textwidth]{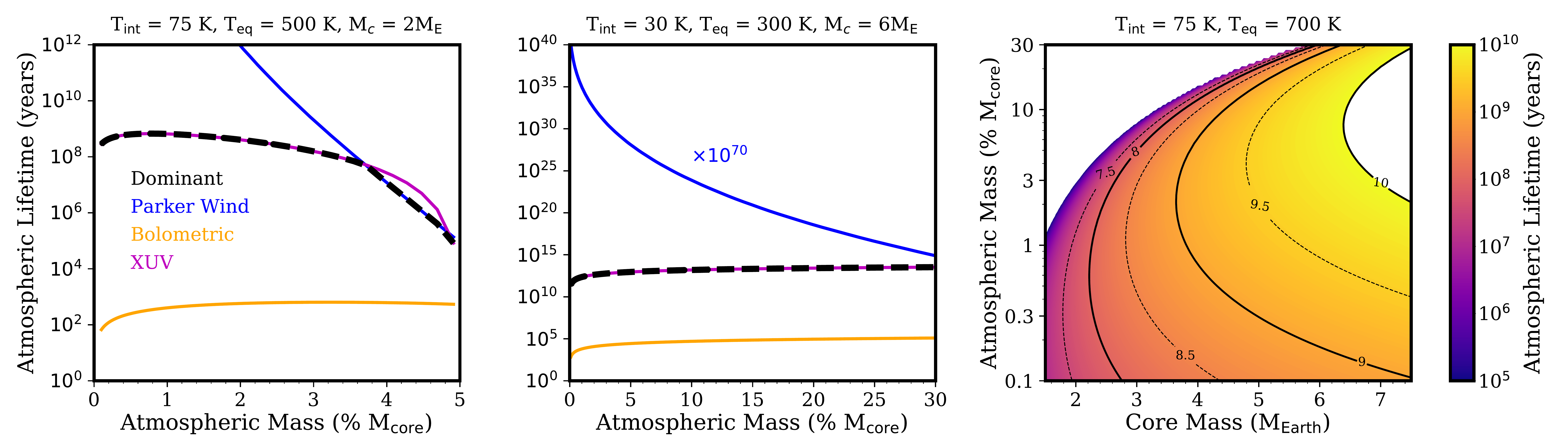}
\caption{The effect of different atmospheric loss mechanisms (isothermal Parker wind: blue; bolometric luminosity-limited: orange; energy-limited photoevaporation from XUV photons: magenta) on the atmospheric lifetime as a function of atmospheric mass for a temperate, young, low mass planet (left) and a cold, old, high mass planet (middle; the Parker wind curve has been reduced by a factor of 10$^{70}$ for clarity). The actual atmospheric lifetime follows the thick, dashed, black curves. (Right) Atmospheric lifetime as a function of core and atmosphere mass for a warm, young planet; the base-10 log of the lifetime is noted in the solid and dotted contour lines. The white space at the top left are parts of parameter space where our model failed to find a RCB (see ${\S}$\ref{sec:radcon}); the white space at the right are models with lifetimes $>$10 Gyr.}
\label{fig:losstime}
\end{figure*}

Note that, while the boil-off has only been hypothesized to affect planets very early in their evolution, before the effects of photoevaporation become significant \citep{owen2016}, we are using it to test whether certain low density model planets can exist. As such, we must consider the boil-off and photoevaporation simultaneously, which leads to the total loss rate $\dot{M}$, defined as

\begin{equation}
\label{eq:mdot}
\dot{M} = max\{\dot{M_e},\dot{M_b}\}.
\end{equation}

\noindent In other words, energy-limited evaporation via photoevaporation is always ongoing, but may be overtaken by boil-off if the planet's density is sufficiently low. 

Energy-limited photoevaporation is the dominant atmospheric loss mechanism for most of the planets in our model grid, allowing for atmospheric lifetimes of $>$1 Gyr (Figure \ref{fig:losstime}). The nonlinear dependence of the atmospheric lifetime as a function of the atmospheric mass fraction, with a peak at a few \%, agrees well with previous works \citep{owen2017,wang2018}, and is a critical phenomenon leading to the radius valley under the photoevaporation mechanism \citep{owen2019}. Planets with atmospheric mass fractions lower than a few \% experience the same loss rate as those with a few \%, since $R_w$, the only variable that can change, does not vary significantly for low atmospheric mass fractions. Photoevaporation still dominates for planets with low mass cores and/or high atmospheric mass fractions, though the relatively weak gravitational potential well of these worlds lead to faster escape and $\tau_a$ $\sim$ 0.1-1 Gyr. Boil-off dominates for the planets with nearly the largest atmospheric mass fractions, when the atmospheric density at the sonic radius becomes sufficiently high; at this point $\tau_a$ drops dramatically and the planet becomes extremely puffy, so much so that $R_w$ becomes large enough that energy-limited photoevaporation dominates for planets with the largest atmospheric mass fractions. For all modeled cases the bolometric luminosity limit is never reached. 

\subsection{Atmospheric Composition}\label{sec:composition}

While we take into account thermochemical equilibrium in our grey opacity calculations to determine the temperature structure \citep{freedman2014}, we assume a simpler complement of gaseous opacity sources in our model atmospheres for computing their transmission spectra. Given a solar metallicity H$_2$/He atmosphere, both Rayleigh scattering from H$_2$ molecules and H$_2$-H$_2$ collision-induced absorption naturally arise. In addition, at the temperatures under consideration at equilibrium, H$_2$O and CH$_4$ are the primary reservoirs for atomic O and C (Figure \ref{fig:tprof}) at pressures $>$1 $\mu$bar and these molecules' spectral features dominate the optical and infrared \citep{lodders2002a,burrows2014}. Therefore, we include them in our model atmospheres assuming mixing ratios equal to those of the atomic species in a solar metallicity, H$_2$/He atmosphere \citep[H$_2$O: 919 ppmv; CH$_4$: 420 ppmv][]{lodders2010}. We do not consider spectral features associated with atomic K and Na, which are common at optical wavelengths of hotter planets, as they should be condensed out in clouds at the temperatures considered here (Figure \ref{fig:tprof}). We also do not consider any other molecular species, as we assume that their opacities are small due to small cross sections and/or abundances. We generate clear and hazy transmission spectra following the methods of \citet{fortney2003,fortney2010}, and include a correction for forward scattering by aerosol particles as formulated by \citet{robinson2017}. We neglect condensation clouds since they form at high pressures (Figure \ref{fig:tprof}).

We treat several physical and chemical processes that affect the mixing ratio profiles of H$_2$O and CH$_4$. For the $T_{eq}$ = 300 K cases, water vapor becomes saturated at 10-100 bars in the atmosphere (Figure \ref{fig:tprof}), where temperatures $\sim$300 K. Water is liquid under such conditions. However, at lower pressures, temperatures could drop to as low as 250 K, yielding water ice. For simplicity, we use the liquid water saturation vapor pressure from \citet{murphy2005} to compute the depletion of water due to this deep cold trap, though we note that the water clouds could be of mixed phase. This should have little effect on our results, however, as the liquid water and water ice saturation vapor pressures are very similar (Figure \ref{fig:tprof}). The H$_2$O mixing ratio is then assumed to be well-mixed above the cold trap until the upper atmosphere, where photochemistry depletes both water and methane (see ${\S}$\ref{sec:hazeprod}). 




\subsection{Photochemistry and Haze Microphysics}\label{sec:hazeprod}

We simulate photochemical haze distributions for our model atmospheres using the Community Aerosol and Radiation Model for Atmospheres (CARMA). CARMA is a 1D bin-scheme aerosol microphysics model that computes vertical and size distributions of aerosol particles by solving the discretized aerosol continuity equation, taking into account aerosol nucleation, condensation, evaporation, coagulation, and transport \citep{turco1979,toon1988,jacobson1994,ackerman1995,gao2014,gao2017a,powell2018,gao2018b,adams2019,powell2019}. We refer the reader to the appendix of \citet{gao2018a} for a complete description of CARMA. 

For this work we rely on a simplified version of the model, where spherical ``seed'' haze particles are generated over a range of pressure levels (see below) and are allowed to grow via coagulation and transported via sedimentation and eddy diffusion \citep{gao2017a,adams2019} while remaining spherical. The atmospheric viscosity is important in setting the sedimentation velocity of aerosol particles, and we use the Sutherland equation for the viscosity of H$_{2}$ gas taken from \citet{white1974} for our model atmospheres,  

\begin{equation}
\label{eq:vis}
\eta(Poise) = 8.76 \times 10^{-5} \left ( \frac{293.85 + 72}{T + 72} \right ) \left ( \frac{T}{293.85} \right )^{1.5}. 
\end{equation}

\noindent Eddy diffusion approximates large scale mass movement in an atmosphere through convective mixing, gravity waves, and circulation, and has been often used in 1D exoplanet atmosphere models \citep[e.g.][]{line2010,kopparapu2012,moses2013a,konopacky2013,miguel2014,hu2014,barman2015,venot2015,tsai2017}. The strength of eddy diffusion is parameterized by $K_{zz}$, the eddy diffusion coefficient, the value of which is uncertain but has been estimated from general circulation models \citep{moses2011,parmentier2013,charnay2015a,zhang2018a,zhang2018b,komacek2019}. For the sub-Neptune GJ 1214b, $K_{zz}$ values at 1 bar of $\sim$10$^7$ cm$^2$ s$^{-1}$ were predicted by \citet{charnay2015a}, with a pressure dependence of $P^{-0.4}$. Here we use a similar but constant value of 10$^8$ cm$^2$ s$^{-1}$ for all of our atmosphere models for simplicity, and we discuss the sensitivity of our results to $K_{zz}$ in ${\S}$\ref{sec:discussion}. In addition to transport of haze by sedimentation and eddy diffusion, we also include an upward wind with velocity $w$ associated with the outflow from atmospheric loss, given by

\begin{equation}
\label{eq:wind}
w(r) = \frac{\dot{M}}{4\pi r^2 \rho(r)}.
\end{equation}

We set the upper boundary condition in CARMA to allow variable fluxes based on the velocity of the particles, while for the lower boundary we set a zero particle number density condition to account for thermal decomposition, though this has little effect on pressures probed in transmission. We allow for 35 bins in the size distribution, with the mass doubling in each successive bin and the largest bin representing $\sim$26 $\mu$m particles. We assume that the seed particles are organic with a mass density of 1 g cm$^{-3}$ and radius 10 nm \citep{adams2019}. Variations in the seed particle radius for spherical particles do not affect the distribution of particles larger than the seed particle, since the coagulation timescale is proportional to the inverse of the particle number density. Thus, mass distributed over a high number density of small seed particles would coagulate faster than the same mass distributed over a low number density of large seed particles, ultimately yielding the same distribution at radii larger than that of the large seed particles. 

The production mechanism of photochemical hazes is highly complex and uncertain. In the atmospheres of Titan and Pluto, methane and nitrogen (and possibly carbon monoxide in the case of Pluto) act as haze parent molecules; their destruction via photolysis and ionization by extreme and far ultraviolet photons from the Sun and, for Titan, energetic particles from Saturn's magnetosphere leads to the formation of radical species that react to form more complex molecules \citep{horst2017,wong2017,luspaykuti2017}. These include polycyclic aromatic hydrocarbons (PAHs), which have been hypothesized as a key step in the production of nm-sized haze particles \citep{wilson2004,trainer2013,yoon2014}. A spectacular array of molecules are produced during this haze formation process with varying numbers of carbon, hydrogen, and nitrogen atoms \citep[e.g.][]{horst2018a}, which have yet to be reproduced by photochemical models. Given the higher temperatures and different volatile contents of exoplanets, we can expect a variety of haze compositions and production pathways, which are beginning to be probed by laboratory studies \citep{horst2018b,he2018,fleury2019}. 

Previous works that rely on detailed photochemical models to derive haze production rates have used parameterizations based on the derived photolysis rates and/or photochemical product abundances. \citet{morley2015} considered a haze forming efficiency that converted some fraction of C$_2$ hydrocarbons, C$_4$H$_2$, and HCN derived from methane and nitrogen photolysis into a haze mass, which were then distributed into haze particles with a lognormal size distribution. \citet{zahnle2016} computed the hydrocarbon haze mass on 51 Eridani b by assuming that all reactions that form C$_4$H$_2$ will eventually go onto forming hazes. \citet{kawashima2018} computed a haze production rate for GJ 1214b by scaling the haze production rate of Titan by the ratio of the Lyman-$\alpha$ flux of GJ 1214 to that received at Titan. \citet{lavvas2017,lavvas2019,kawashima2019a,kawashima2019b} all used similar approaches, where the haze production rate is set equal to the photolysis rates of methane, nitrogen, and major hydrocarbon and nitrile species, reduced by an efficiency factor. 


In this work we use a simplified photochemical scheme so that we can expand our coverage of parameter space and get a better understanding of some basic controls on haze production on sub-Neptunes and super-puffs. We assume that haze stems solely from photolysis of methane, the main carbon-carrier in this temperature regime, by Lyman-$\alpha$ radiation from the host star, for which the only competing absorber is water vapor. We discuss complications to this picture in ${\S}$\ref{sec:photocons}. We do not consider nitrogen since its abundances are lower than carbon by a factor of 3 in a solar metallicity atmosphere \citep{lodders2010} and nitrogen photolysis requires much higher energy photons, though nitrogen incorporation into organic hazes is an important process in solar system atmospheres \citep[e.g.][]{vuitton2007}. We also ignore the impact of sulfur hazes formed from H$_2$S photochemistry, since they likely form at higher pressures than the hydrocarbon hazes \citep{zahnle2016,gao2017b}. 

\begin{figure*}[hbt!]
\centering
\includegraphics[width=0.8 \textwidth]{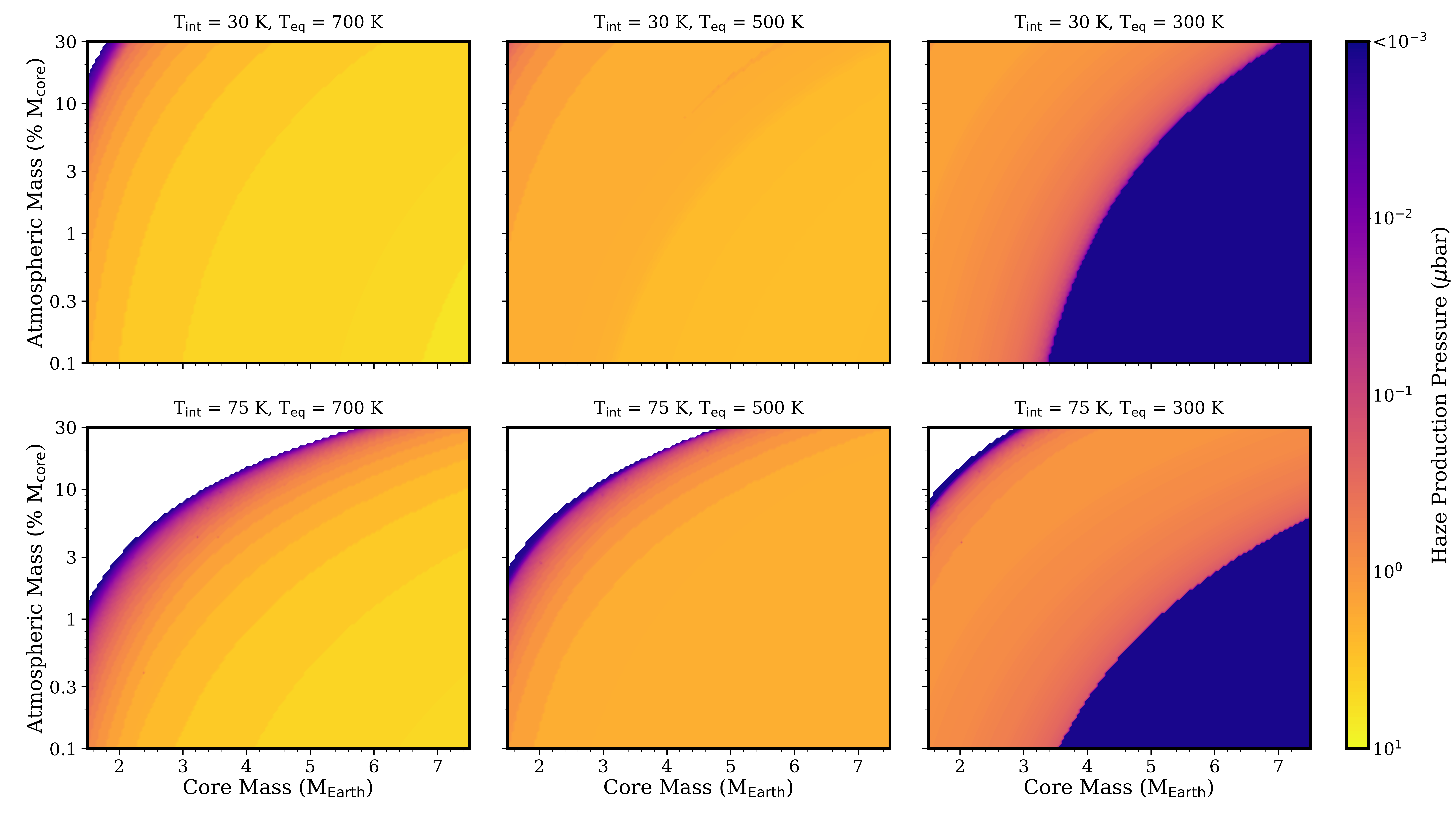}
\caption{The pressure at which the methane photolysis timescale equals the transport timescale for all considered cases. The white space at the top left are parts of parameter space where our model failed to find a RCB (see ${\S}$\ref{sec:radcon}).}
\label{fig:hazeprod}
\end{figure*}

Starting with initial water and methane abundance profiles computed in ${\S}$\ref{sec:composition}, the methane and water photolysis rates, $\mathcal{L}^i_{C\gamma}$ and $\mathcal{L}^i_{H\gamma}$ at altitude level $i$ are 

\begin{equation}
\label{eq:lfluxch4}
\mathcal{L}^i_{C\gamma} = \frac{I_{Ly\alpha}}{4a^2 \delta_i}\left ( 1 - e^{-\tau^i_{\alpha}}\right ) \frac{\tau^i_{\alpha(CH_4)}}{\tau^i_{\alpha}}
\end{equation}

\begin{equation}
\label{eq:lfluxh2o}
\mathcal{L}^i_{H\gamma} = \frac{I_{Ly\alpha}}{4a^2 \delta_i}\left ( 1 - e^{-\tau^i_{\alpha}}\right ) \frac{\tau^i_{\alpha(H_2O)}}{\tau^i_{\alpha}}
\end{equation}

\noindent where $I_{Ly\alpha}$ = 3.7 $\times$ 10$^{11}$ cm$^{-2}$ s$^{-1}$ is the Lyman-$\alpha$ flux at 1 AU from the Sun \citep{krasnopolsky2004}, $a$ is the semi-major axis of the planet in AU, $\delta_i$ is the thickness of level $i$, the factor of 4 accounts for the global-averaging of incident Lyman-$\alpha$, and $\tau^i_{\alpha}$ is the nadir Lyman-$\alpha$ optical depth at altitude level $i$ defined as 

\begin{multline}
\label{eq:taualpha}
\tau^i_{\alpha} = \tau^i_{\alpha(CH_4)} + \tau^i_{\alpha(H_2O)} = \\ 
N_i\delta_i \left ( C_{CH_4}f^i_{CH_4} + C_{H_2O}f^i_{H_2O} \right )
\end{multline}

\noindent where $N_i$ is the total number density of the atmosphere at level $i$ and $f^i_x$ and $C_x$ are the mixing ratio at level $i$ and Lyman-$\alpha$ cross section of molecule \textit{x}, respectively. For simplicity, we set C$_{CH_4}$ and C$_{H_2O}$ to be the average of the two values \citep[1.8 $\times$ 10$^{-17}$ cm$^2$ for methane and 1.53 $\times$ 10$^{-17}$ cm$^2$ for water;][]{heays2017}. 

The loss of methane (water) to photolysis is balanced by resupply via upward transport of gases from depth by mixing, which we parameterize using $K_{zz}$ for eddy and molecular diffusion, and the outflow wind from atmospheric escape. We define a ``base'' for the photolysis region of methane (water), $r_{C\gamma}$ ($r_{H\gamma}$), as the altitude level at which the methane (water) photolysis timescale, defined as the methane (water) concentration divided by $\mathcal{L}^i_{C\gamma}$ ($\mathcal{L}^i_{H\gamma}$), equals the transport timescale $\tau_{trans}$, defined as 

\begin{equation}
\label{eq:tautrans}
\tau_{trans} = \left ( \tau^{-1}_{wind} + \tau^{-1}_{eddy} \right )^{-1}
\end{equation}

\noindent with

\begin{equation}
\label{eq:tauwind}
\tau_{wind} = \frac{H}{w}
\end{equation}

\begin{equation}
\label{eq:taueddy}
\tau_{eddy} = \frac{H^2}{K_{zz}}
\end{equation}

\noindent as the wind transport and mixing timescales, respectively, where $H$ is the scale height. From $r_{C\gamma}$ and $r_{H\gamma}$, we assume that the methane and water mixing ratios fall off linearly with increasing altitude with slopes $\beta_C$ and $\beta_H$, respectively, 

\begin{equation}
\label{eq:mrfallch4}
f^i_{CH_4} = f^{r_{C\gamma}}_{CH_4} - \beta_C \left ( r_i - r_{C\gamma} \right )
\end{equation}

\begin{equation}
\label{eq:mrfallh2o}
f^i_{H_2O} = f^{r_{H\gamma}}_{H_2O} - \beta_H \left ( r_i - r_{H\gamma} \right )
\end{equation}

\noindent where $r_i$ is the altitude at level $i$ and $ f^{r_{C\gamma}}_{CH_4}$ ($ f^{r_{H\gamma}}_{H_2O}$) is the methane (water) mixing ratio at $r_{C\gamma}$ ($r_{H\gamma}$). We can then write a continuity equation that solves for each $\beta$ in spherical coordinates to take into account extended atmospheres, e.g. 

\begin{equation}
\label{eq:continuity}
\left. \frac{1}{r^2} \frac{\partial}{\partial r} \left ( -r^2 N K_{zz}\frac{\partial f}{\partial r} + r^2fNw \right ) \right\rvert_{r=r_i} = -\mathcal{L}^i_{C\gamma}
\end{equation}

\noindent for methane. Taking the integral from $r_{C\gamma}$ to the top of the model atmosphere at 1 nbar, $r_{top}$, of both sides of Eq. \ref{eq:continuity} and noting that $\partial f / \partial r = -\beta_C$ (Eq. \ref{eq:mrfallch4}) and that $r^2Nw$ is constant (Eq. \ref{eq:wind}), we find

\begin{equation}
\label{eq:slope}
\beta_C = \frac{\int_{r_{C\gamma}}^{r_{top}} \mathcal{L}_{C\gamma}(r')r'^2dr}{\mathcal{A}\left (r_{top} - r_{C\gamma} \right ) + K_{zz} \left ( \mathcal{R}^2_{C\gamma}N_{C\gamma} - r^2_{top}N_{top} \right )}
\end{equation}

\noindent where $N_{C\gamma}$ and $N_{top}$ are the atmospheric number density at $r_{C\gamma}$ and $r_{top}$, respectively, and $\mathcal{A}$ = $\dot{M}A/4\pi\mu$ is a constant, with $A$ as Avogadro's number. Eqs. \ref{eq:lfluxch4}-\ref{eq:slope}, including versions of Eqs. \ref{eq:continuity}-\ref{eq:slope} for water, are iterated until $r_{C\gamma}$ stops changing by more than 10 ppm between iterations. An example of converged mixing ratio and photolysis rate profiles is shown in Figure \ref{fig:tprof}. 

To obtain the haze production rate profile, we multiply the methane photolysis rate profile by a haze efficiency factor $\epsilon_h$, which we set to 10\%; this results in column haze production rates of 3 $\times$ 10$^{-12}$, 8 $\times$ 10$^{-13}$, and $\sim$3 $\times$ 10$^{-13}$ g cm$^{-2}$ s$^{-1}$ for the $T_{eq}$ = 700, 500, and 300 K cases, respectively, which are distributed into haze seed particles with radii of 10 nm. The approximate value for the 300 K cases are due to varying water vapor mixing ratios from the deep cold trap. These values and the overall shape and location of the methane photolysis profile are consistent with those of previous modeling studies \citep[e.g.][]{lavvas2017,kawashima2019b}. We evaluate the sensitivity of our results to $\epsilon_h$ in ${\S}$\ref{sec:effvar}.

Our parameterization of the haze production rate leads to variations in the pressure at $r_{C\gamma}$ (Figure \ref{fig:hazeprod}). As atmospheric mass increases and local gravity decreases due to decreasing core mass and/or increasing temperatures creating more extended atmospheres, the opacity above any given pressure level increases, pushing $r_{C\gamma}$ to lower pressures. Conversely, we also find increasing $r_{C\gamma}$ for lower atmospheric masses and higher local gravities, particularly for the coolest model cases. This is due to our assumption of a constant $K_{zz}$ for all models, such that $\tau_{eddy}$ becomes small for small scale height objects and mixing is able to rapidly replenish methane in the upper atmosphere. As $K_{zz}$ is likely to vary with temperature and scale height, we do not consider this effect to be real, though how $K_{zz}$ should vary with these parameters could be complex. 

For the cases where $r_{C\gamma}$ $>$ $r_{top}$ because of high $K_{zz}$, our algorithm is capable of generating a photolysis rate profile, though it will not deplete methane at pressures $>$ 1 nbar. On the other hand, if the atmosphere is sufficiently extended such that it is opaque to Lyman-$\alpha$ at pressures $\geq$1 nbar, then we assume a downward flux of haze seed particles at the upper boundary of the model atmosphere equal to the column haze production rate. 

\begin{figure*}[hbt!]
\centering
\includegraphics[width=1.0 \textwidth]{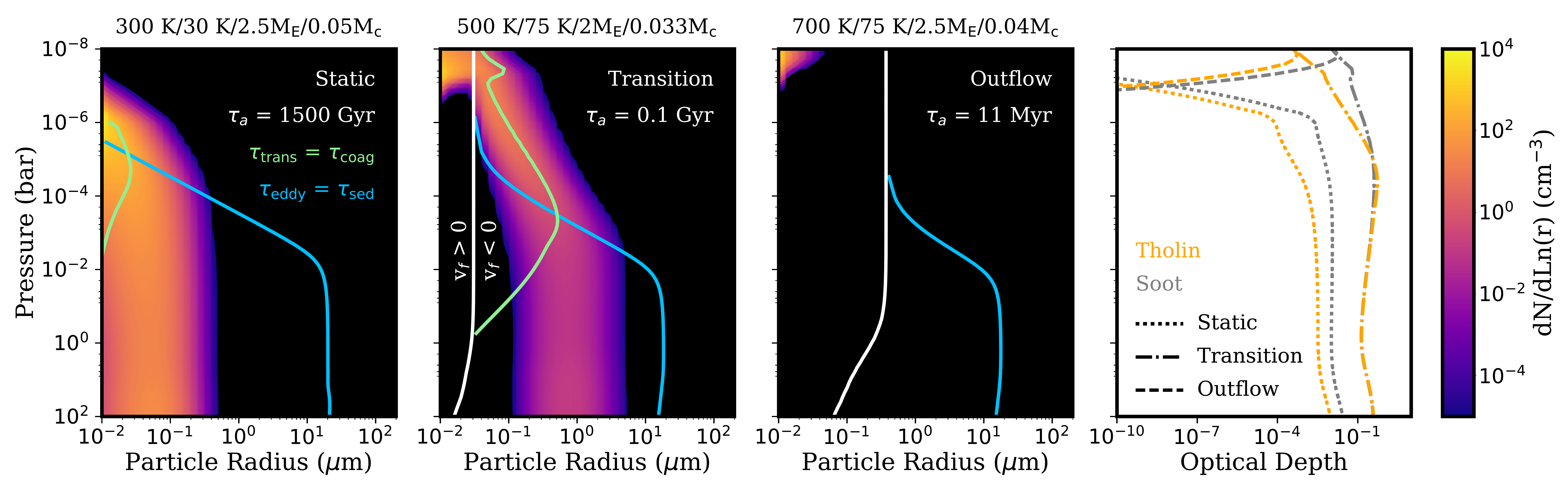}
\caption{Haze distributions for different $T_{eq}$/$T_{int}$/$M_c$/$M_a$ cases (left, middle-left, middle-right) and associated optical depth profiles in the Kepler bandpass (right), assuming both soot (gray) and tholin (orange) refractive indices. The different cases were chosen to illustrate the ``static'', ``transition'', and ``outflow'' types of haze distributions, and the associated atmospheric lifetimes are given as $\tau_a$. The blue curve marks where the sedimentation timescale equals the eddy diffusion timescale, while the green curve marks where the transport timescale equals the coagulation timescale. The white curve marks where the particle velocity is zero due to balance between the sedimentation and outflow wind velocities. }
\label{fig:hazedist}
\end{figure*}

We treat the haze particles as Mie spheres \citep{bohren2008} in calculating their optical properties. As the haze composition is uncertain, we consider the refractive indices of both the Titan haze analog, tholins \citep{khare1984}, as well as soots  \citep{morley2015,lavvas2017}. As soots are more absorbing than tholins in our wavelengths of interest \citep[$\sim$0.3-5 $\mu$m;][]{adams2019}, rather than assigning their refractive indices to their compositions, we treat tholins as representing scattering hazes, and soots as representing absorbing hazes. 

\section{Results}\label{sec:results}

\subsection{Static, Outflow, and Transition Hazes}\label{sec:hazetypes}

The inclusion of an outflow velocity to the dynamics of the haze leads to a continuum of haze distributions as a function of atmospheric loss rate (Figure \ref{fig:hazedist}). For atmospheres with low loss rates, and thus negligible outflow velocities (the ``static'' case), the haze is controlled by a balance of particle coagulation, sedimentation, and diffusion timescales, $\tau_{coag}$, $\tau_{sed}$, and $\tau_{eddy}$, respectively, with 

\begin{equation}
\label{eq:taucoag}
\tau^j_{coag} = \frac{1}{\sum_i \mathcal{K}_{ij} n_i}
\end{equation}

\begin{equation}
\label{eq:taused}
\tau_{sed} = \frac{H}{v_f}
\end{equation}

\noindent where $\tau^j_{coag}$ refers to the coagulation timescale of particles in bin $j$, with the summation in the denominator indicating coagulation with particles in all bins $i$ (including $j$), each with number density $n_i$; $\mathcal{K}_{ij}$ is the coagulation kernel between bins $i$ and $j$; and $v_f$ is the fall velocity \citep[see Appendix of][]{gao2018a}.  

These static hazes have characteristic distributions: For our fixed $K_{zz}$, transport of hazes at the pressures where they are produced is dominated by sedimentation, and thus the haze particles grow as much as they can via coagulation before they fall. In Figure \ref{fig:hazedist}, this is shown in the left plot, where the peak of the haze distribution follows the green curve at pressures $<$ 10 $\mu$bar, indicating where  $\tau_{trans}$ = $\tau_{coag}$. At higher pressures, where sedimentation velocities decrease from rising atmospheric density, transport by eddy diffusion dominates and further haze growth is quenched. This is analogous to quenching of chemical reactions in warm exoplanet atmospheres \citep[e.g.][]{moses2013b}. Consequently, the haze distribution varies little from the quenching point, where $\tau_{coag}$ = $\tau_{sed}$ = $\tau_{eddy}$, to deeper pressures, assuming the haze particles remain thermally stable. At the opposite extreme, where the atmospheric loss rate is high and the planet cannot survive for more than a few Myr (the ``outflow'' case; middle-right of Figure \ref{fig:hazedist}), very little haze forms since the outward wind speed carries away any haze seed particles that form from photo- and ion chemistry before they have time to grow by coagulation.  

The ``transition'' case, in between the two extremes, offers the most interesting scenario: The outflow wind speed is high enough to entrain the smallest of seed particles, but not sufficiently high to prevent coagulation. As particles grow, their sedimentation velocities increase and their upward speed reduces, resulting in $\tau_{trans}\to\infty$ and runaway growth via coagulation until a particle size is reached such that sedimentation begins to dominate. Subsequent evolution of the haze is similar to the static case, except all particles are larger since their sedimentation velocity is reduced by the outflow. In addition, the haze seed particles are produced at lower pressures compared to the static case due to higher Lyman-$\alpha$ opacities in the puffier transition case atmospheres. 

As a result of the large particles at low pressures and the outward wind, the transition case features the largest haze opacity at all pressures (right of Figure \ref{fig:hazedist}). In comparison, the static case does not have large particles at low pressures and the haze is produced at higher pressures due to lower Lyman-$\alpha$ opacities, and so it has lower optical depths; the outflow case does have haze production at low pressures, but the particles remain small and are not transported to higher pressures, resulting in high optical depths at only low pressures. The effect of the particle size is seen in the different optical depths between tholin hazes and soot hazes. The extinction coefficient of particles with radii comparable or larger than the wavelength of interest will converge to $\sim$2 regardless of refractive index values; this is seen for the transition case, where the optical depth profiles of soots and tholins are identical at pressures $>$10 $\mu$bar. In contrast, the other two cases show large differences between soots and tholins, with the less absorptive tholins exhibiting lower optical depth, indicative of particles smaller than the wavelength of interest. 

We can place the transition case in the context of atmospheric lifetimes by noting that the outflow wind speed must be similar to the sedimentation velocity of haze seed particles at the pressure level of haze formation, 

\begin{equation}
\label{eq:fall_eq_wind}
\frac{2}{9}\frac{\rho_p g R_p^2}{\eta}\mathcal{B} = \frac{\dot{M}_h}{4\pi r^2 \rho}
\end{equation}

\begin{figure}[hbt!]
\centering
\includegraphics[width=0.45 \textwidth]{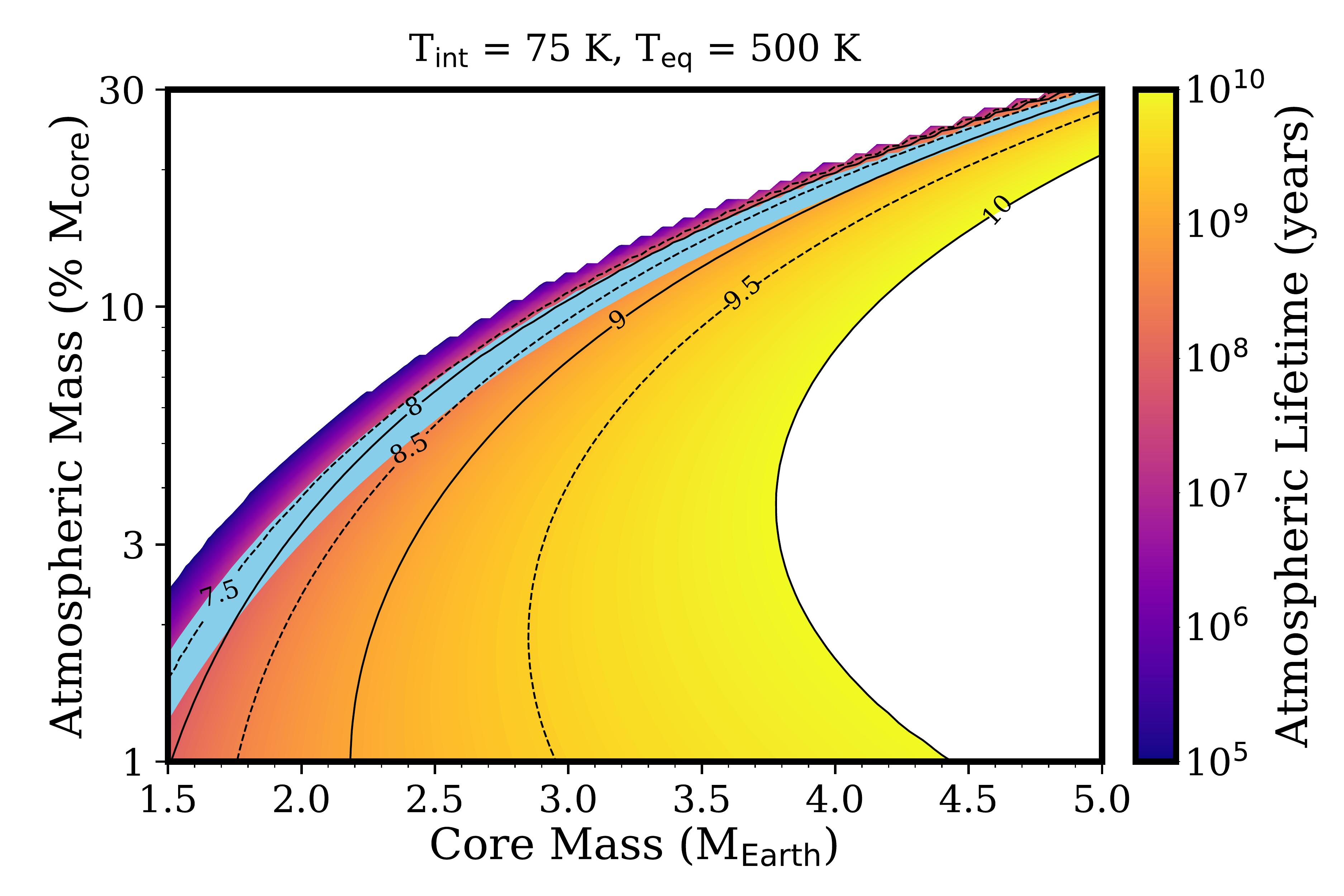}
\caption{Atmospheric lifetime as a function of atmospheric mass and core mass for temperate, young planets, with the region of parameter space where transition hazes occur marked in light blue, defined using R$_p$ = 10-100 nm and T$_h$ = 400 K.}
\label{fig:transition}
\end{figure}

\begin{figure*}[hbt!]
\centering
\includegraphics[width=0.7 \textwidth]{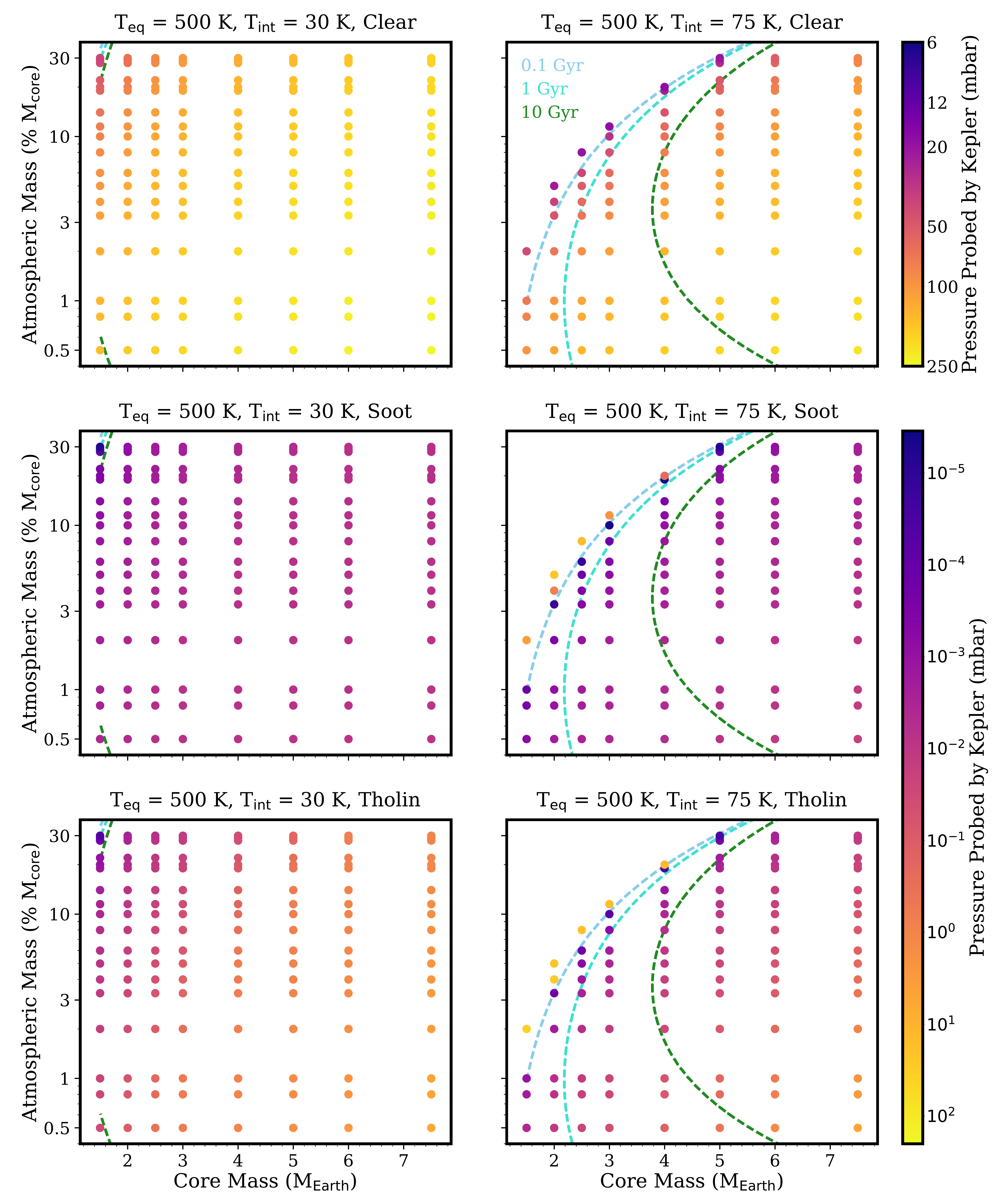}
\caption{The highest atmospheric pressure probed in the Kepler bandpass for clear (top) and hazy (soot: middle; tholin: bottom) objects with $T_{eq}$ = 500 K. Note the different colorbars for the clear and hazy cases. Contours of atmospheric lifetimes are shown for 0.1 Gyr (light blue), 1 Gyr (cyan), and 10 Gyr (green). Results for other $T_{eq}$ cases can be found in Appendix \ref{sec:appprerad}}
\label{fig:teq500pressure}
\end{figure*}

\noindent where we have set the Stokes fall velocity to the wind velocity from Eq. \ref{eq:wind}. $\dot{M}_h$ is the atmospheric loss rate associated with the existence of a transition haze, $\rho_p$ and $R_p$ are the mass density and radius of the haze particle, $\eta$ is the dynamic viscosity of the atmosphere, and $\mathcal{B}$ is the Cunningham slip correction factor, given by 

\begin{equation}
\label{eq:slip}
\mathcal{B} = 1 + 1.246 {\rm Kn }+ 0.42 {\rm Kn} e^{-0.87/{\rm Kn}}
\end{equation}

\noindent where Kn is the Knudsen number, equal to the ratio of the mean free path $l$ of the atmosphere to R$_p$, with,

\begin{equation}
\label{eq:mfp}
l = \frac{2\eta}{\rho}\sqrt{\frac{\pi\mu}{2RT_h}}
\end{equation}

\noindent where T$_h$ is the temperature of the isothermal region in the radiative zone. At the pressure level of haze formation ($\leq$1 $\mu$bar), Kn$\sim$10$^8$-10$^7$ for particles with R$_p$ = 10-100 nm, respectively, and so $\mathcal{B} \sim 1.662$Kn. Including this in Eq. \ref{eq:fall_eq_wind} and noting that g = $GM_c/r^2$ and $\dot{M}_h$ = M$_a / \tau^h_a$, where $\tau^h_a$ is the corresponding atmospheric lifetime, we find

\begin{multline}
\label{eq:transitionmdot}
\tau^h_a = \frac{\overline{M}_a}{11.6 G \rho_p R_p}\sqrt{\frac{RT_h}{\mu}}  
\sim \\
(0.25\, {\rm Gyr}) \left (\frac{\overline{M}_a}{0.05} \right ) \left (\frac{\rho_p}{1\, {\rm g\, cm}^{-3}} \right )^{-1} \left (\frac{R_p}{10\, {\rm nm}} \right )^{-1} \\
\left (\frac{T_h}{400\, {\rm K}} \right )^{1/2} \left (\frac{\mu}{2.3\, {\rm g\, mol}^{-1}} \right )^{-1/2} 
\end{multline}

\noindent where $\overline{M}_a$ = $M_a$/$M_c$. A transition haze forms on a planet when $\tau^h_a$ = $\tau_a$, and thus temperate sub-Neptunes with ages $\sim$0.1-1 Gyr, such as those in the Kepler-51 system, are the perfect candidates to host such hazes (Figure \ref{fig:transition}).  

\subsection{Optical Transit Pressures and Radii}


The atmospheric pressures probed by transits in the \textit{Kepler} bandpass in clear, low mass planets decreases with decreasing core mass and increasing atmosphere mass fraction, ranging from a few hundred mbar to $\leq$10 mbar (Figure \ref{fig:teq500pressure}). This is caused by increased atmospheric opacity stemming from decreasing local gravity when the core mass is decreased or the atmospheric mass is increased, similar to how the Lyman-$\alpha$ photosphere varies with planetary parameters (Figure \ref{fig:hazeprod}). 

\begin{figure*}[hbt!]
\centering
\includegraphics[width=0.7 \textwidth]{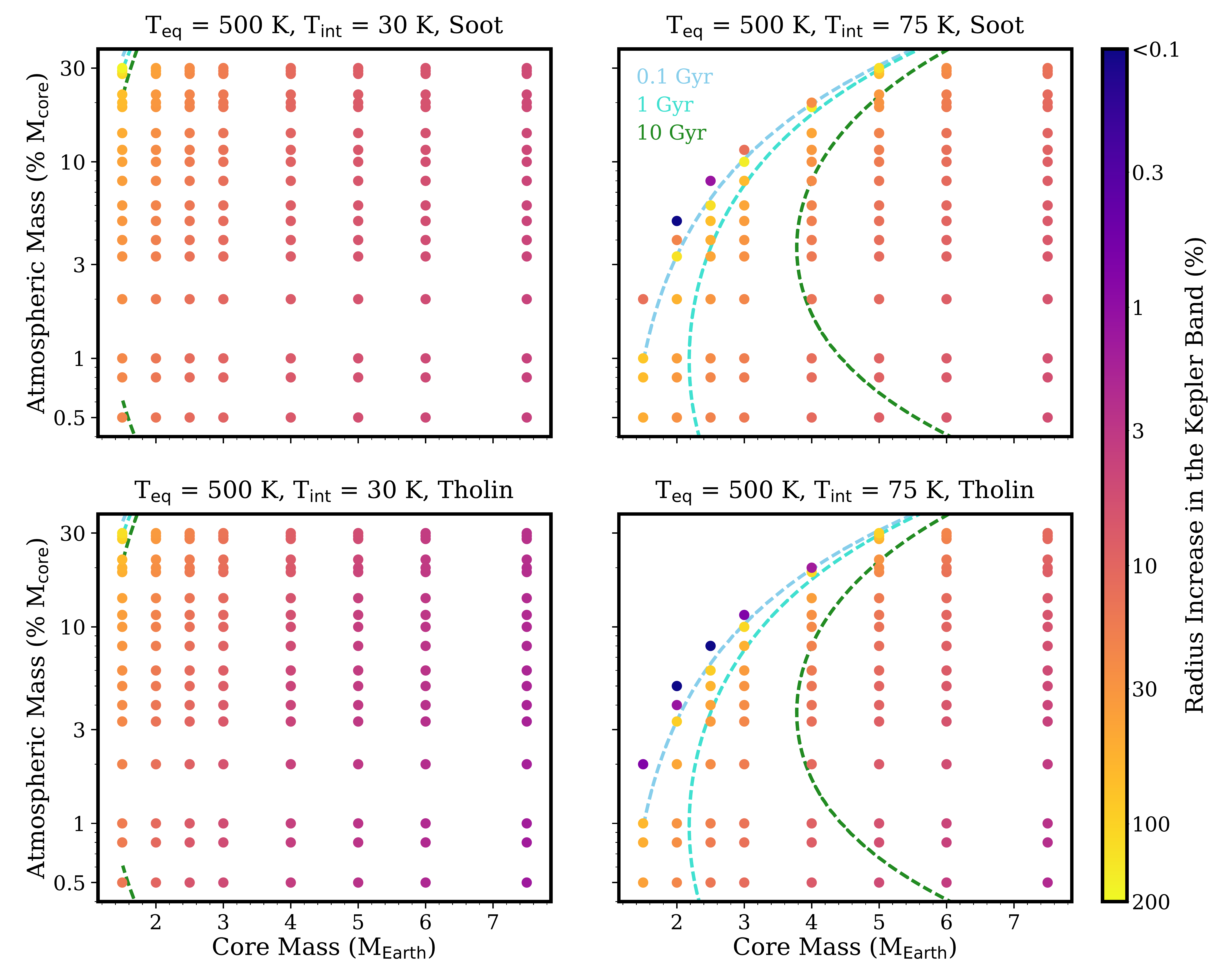}
\caption{The increase in planet radius due to a high altitude photochemical haze (soot: top; tholin: bottom) in the Kepler bandpass for $T_{eq}$ = 500 K objects, defined as the ratio of planet radius with a haze to that without a haze, minus 1. Contours of atmospheric lifetimes are shown for 0.1 Gyr (light blue), 1 Gyr (cyan), and 10 Gyr (green). Results for other $T_{eq}$ cases can be found in Appendix \ref{sec:appprerad}. }
\label{fig:teq500raddif}
\end{figure*}

In contrast, the inclusion of opacity from photochemical hazes drastically reduces the atmospheric pressures probed in optical transmission, with the magnitude of the reduction depending on whether the planet hosts static, transition, or outflow hazes. Atmospheres that host static hazes can only be probed to 0.01-10 mbar depending on whether the hazes are absorbing (lower pressures probed) or scattering (higher pressures probed). The pressures probed in transmission gradually reduces further as the outflow wind increases in intensity: optical transits can only probe to 0.01-0.1 $\mu$bar in atmospheres hosting transition hazes, consistent with that required to explain the near-IR transmission spectra of super-puffs \citep{libbyroberts2019}. This is caused by increasing haze opacity in the upper atmosphere with increasing outflow wind speeds (${\S}$\ref{sec:hazetypes}). This trend reverses at the highest wind speeds, however, due to the entrainment of haze in the wind and quenching of haze particle growth (i.e. outflow hazes), leading to pressures probed in optical transits similar to those of the corresponding clear atmosphere cases. 

The reduction in pressures probed by optical transits caused by high altitude hazes directly translates to increases in the transit radii observed by \textit{Kepler} (Figures \ref{fig:teq500raddif}). Radius enhancement varies smoothly from a few \% for static haze-hosting planets to nearly 200\% for transition haze-hosting planets, while outflow haze-hosting planets see little change in radius due to optically thin hazes. The extraordinary increase in transit radii of transition haze-hosting planets is caused by (1) moderate outflow winds increasing the altitude of haze formation, (2) the lofting of large haze particles to low pressures without quenching coagulation, and (3) large atmospheric scale heights stemming from a large atmospheric mass fraction. 


Combining our results for radius enhancement with our delineation of where in parameter space transition hazes can form, we find that objects that are young ($\sim$0.1-1 Gyr), warm (T$_{eq}$ $\geq$ 500 K), and low mass ($M_c$ $<$ 4M$_{\Earth}$) should experience the most radius enhancement due to hazes, with the hazy atmosphere radius nearly three times that of the clear atmosphere radius. Higher mass objects can also exhibit transition hazes, but they would require atmospheric mass fractions greater than a few \% and/or higher $T_{int}$ and $T_{eq}$.

\begin{figure*}[hbt!]
\centering
\includegraphics[width=0.7 \textwidth]{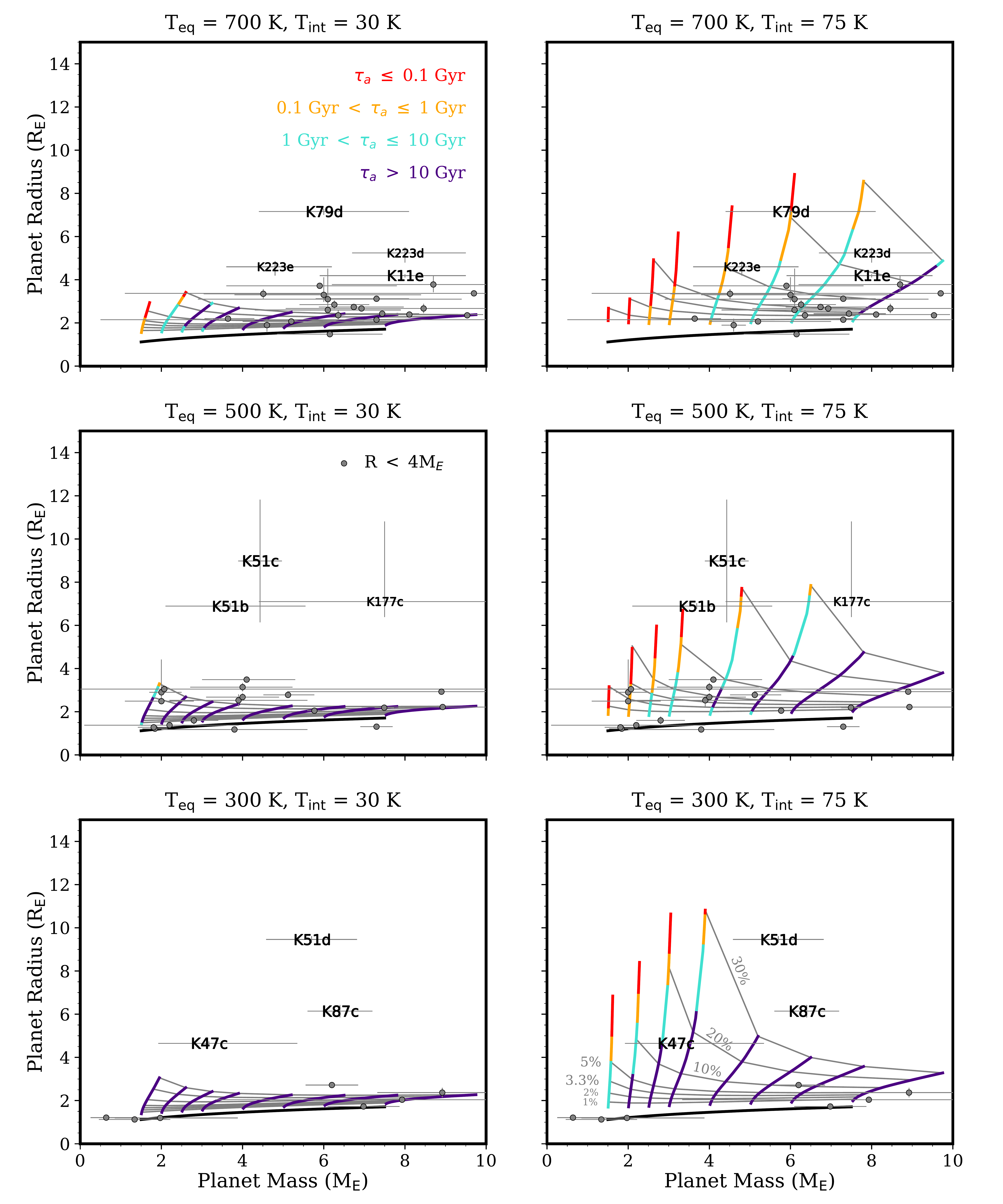}
\caption{Mass-radius diagrams for sub-Neptunes and super puffs with clear atmospheres and various $T_{int}$ and $T_{eq}$ values. Each multi-colored curve indicates the trend in mass-radius with increasing atmospheric mass, marked in \% of core mass by the gray lines. The colors along each mass-radius trend designate planets with atmospheric lifetimes $<$0.1 Gyr (red), between 0.1 and 1 Gyr (orange), between 1 and 10 Gyr (cyan), and $>$10 Gyr (indigo). The mass-radius relationship for the rocky core is shown in black \citep{zeng2019}. Observed masses and radii of  sub-Neptunes are shown by the gray points, with those in the top plots having equilibrium temperatures 600 K $\leq$ $T_{eq}$ $<$ 800 K; those in the middle plots with 400 K $\leq$ $T_{eq}$ $<$ 600 K; and those in the bottom plots with $T_{eq}$ $<$ 400 K. Super-puffs are shown by their shortened names (K = Kepler). Only planets with $T_{eq}$ $<$ 800 K, masses $<$ 10M$_{\Earth}$, and radii $>$ 1.5R$_{\Earth}$ are presented.}
\label{fig:clearmr}
\end{figure*}

\begin{figure*}[hbt!]
\centering
\includegraphics[width=0.7 \textwidth]{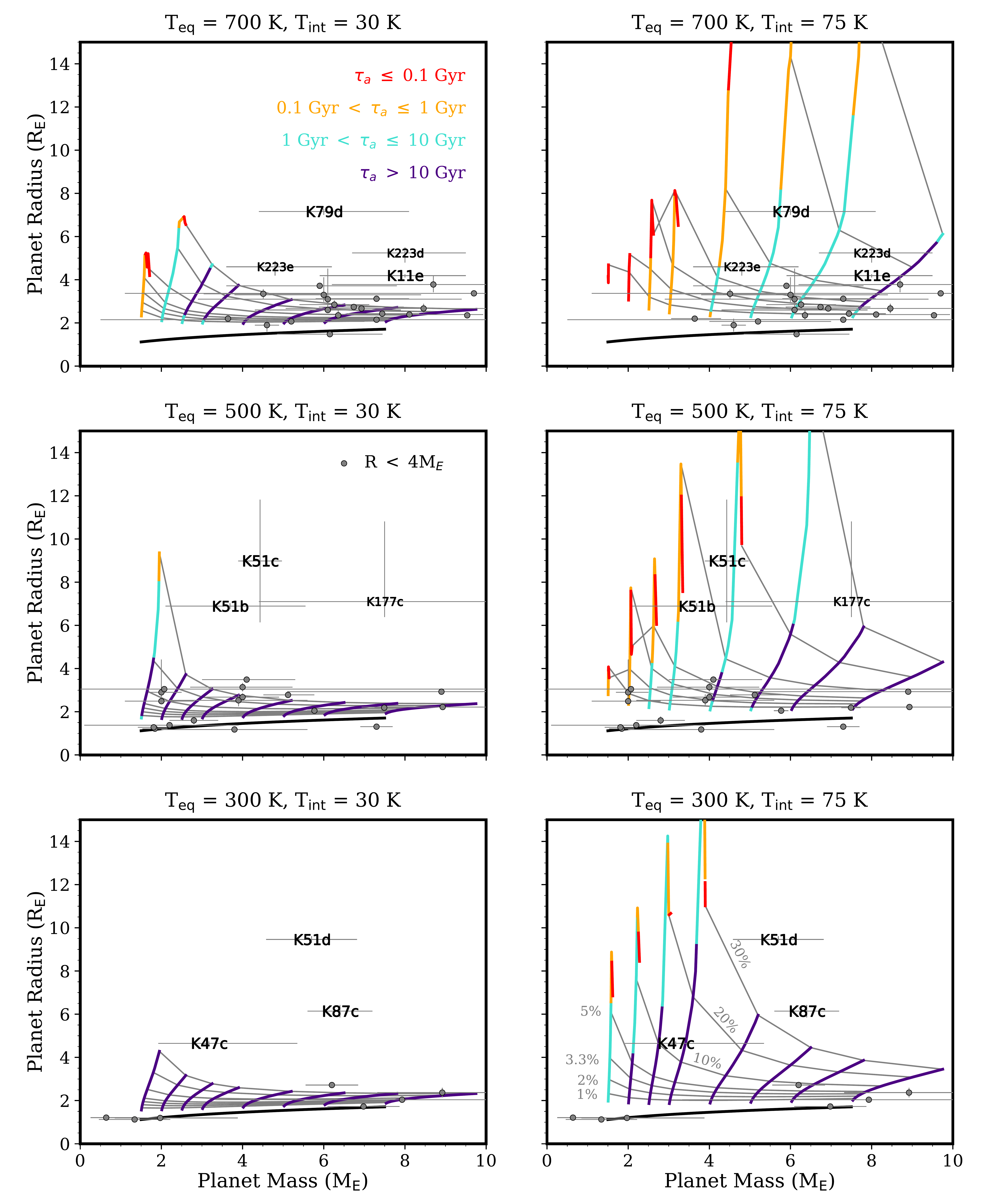}
\caption{Same as Figure \ref{fig:clearmr}, but for planets with soot hazes. Results for tholin hazes can be found in Appendix \ref{sec:appthomassrad}}
\label{fig:sootmr}
\end{figure*}

\subsection{A Hazy Mass-Radius Diagram}

We construct mass-radius diagrams for low mass planets that account for their atmospheric lifetimes and the effects of high altitude photochemical hazes (Figures \ref{fig:clearmr}-\ref{fig:sootmr}). Our clear atmosphere results are similar to those of previous works that considered the impact of adding a H$_2$/He atmosphere to a rocky core. For example, we are able to reproduce the rapid increase in planet radius with the addition of only small abundances of gas, the increase in planet radius with temperature for a fixed gas mass fraction, and the increase in planet radius along fixed gas mass fraction contours towards both small and large core masses \citep{seager2007,mordasini2012,zeng2019}. \citet{rogers2011} evaluated the impact of atmospheric loss via energy limited escape on the gas mass fraction of planets with a given radius, but did not map the atmospheric lifetimes directly onto their mass-radius diagram. They also considered Roche lobe overflow, which we do not, and found that it should not greatly affect planets with $T_{eq}$ $<$ 1000 K around sun-like stars. 

For both clear and hazy cases, increasing $T_{eq}$ reduces the lifetimes and maximum radii achievable for lower mass worlds ($M_c$ $<$ 4M$_{\Earth}$) due to increasing atmospheric escape, while the opposite trend exists for higher mass planets, as their atmospheres are not readily lost. The hazy cases exhibit an interesting phenomenon where the fixed core mass tracks turn downwards at short atmospheric lifetimes. This is due to the reduced haze opacity of these cases, which host outflow hazes, allowing optical transits to probe higher pressures and thus smaller radii. Lower $T_{int}$, which we use to approximate older planets, leads to significantly reduced radii, since the convective region has lower entropy than in high $T_{int}$ cases. However, direct comparisons in age are difficult since (1) only a few of these planets' host stars have measured ages, and often with conflicted findings and/or large uncertainties \citep[e.g.][]{masuda2014}, and (2) $T_{int}$ varies with planet mass for a fixed age, with higher mass planets possessing $T_{int}$ $>$ 30 K even at an age of 5 Gyr \citep{lopez2014}. 

\begin{deluxetable*}{lccccc}
\tablecolumns{6}
\tablecaption{Super-puff candidates. \label{table:youngplanets}}
\tablehead{
\colhead{Planet} & \colhead{Mass Constraint} & \colhead{Radius (R$_{\Earth}$)} & \colhead{$T_{eq}$ (K)} & \colhead{Age (Myr)} & \colhead{References}}
\startdata
K2-33 b & $<$3.6M$_{J}$ & 5.76 $^{+0.62}_{-0.58}$ & 850$^{+50}_{-50}$  & 9.3 $^{+1.1}_{-1.30}$ & \citet{mann2016,david2016} \\
V1298 Tau b	& $<$120M$_{\Earth}$ & 10.27 $^{+0.58}_{-0.53}$ & 677 $^{+22}_{-22}$ & 23$^{+4}_{-4}$ & \citet{david2019a,david2019b} \\
V1298 Tau c	& $<$28M$_{\Earth}$ & 5.59 $^{+0.36}_{-0.32}$ & 968 $^{+31}_{-31}$& 23$^{+4}_{-4}$ & \citet{david2019b} \\
V1298 Tau d	& $<$28M$_{\Earth}$ &  6.41 $^{+0.45}_{-0.40}$ & 847 $^{+27}_{-27}$& 23$^{+4}_{-4}$ & \citet{david2019b} \\
V1298 Tau e	& \nodata & 8.74 $^{+0.84}_{-0.72}$ & 492 $^{+66}_{-104}$& 23$^{+4}_{-4}$ & \citet{david2019b} \\
DS Tuc A b & $<$1.3M$_{ J}$ & 5.70$^{+0.17}_{-0.17}$ & 850  & 45$^{+4}_{-4}$ & \citet{newton2019,benatti2019} \\
\enddata
\end{deluxetable*}

The inclusion of high altitude photochemical hazes helps explain the observed masses, radii, and ages of several super-puffs. The most extreme examples are Kepler-51b and c, which possess far shorter lifetimes than the inferred age of their host star if clear atmospheres were assumed. As pointed out in ${\S}$\ref{sec:hazetypes}, however, these planets should possess transition hazes that drastically enhance the observed transit radius. This is indeed the case, as including the effect of soot hazes increases their lifetimes to $>$0.1 Gyr (assuming $T_{int}$ = 75 K), in line with the system's age. For Kepler-51b, adding hazes also results in an atmospheric mass fraction of $\leq$10\%, similar to that of the large radii population of sub-Neptunes \citep{owen2017}. This is in contrast to the 16.9\% computed by \citet{lopez2014} without taking into account aerosols (see ${\S}$\ref{sec:kepler51b}). For Kepler-51c, adding hazes lead to an atmospheric mass of $\sim$0.15$M_c$, though lower gas masses are allowed if it possesses a higher $T_{int}$, which is possible given its higher mass. 

The lifetimes of Kepler-79d and Kepler-223e can also be brought to values more consistent with their inferred ages, though relatively high $T_{int}$'s are required given said ages, which could be sustained through obliquity tides \citep{millholland2019}. Including hazes also decreases their atmospheric masses by $\sim$30\% compared to the clear atmosphere cases, with a hazy Kepler-223e needing only $\sim$6\% gas fraction. On the other hand, if their $T_{int}$'s are lower due to their age, then they may require gas masses $>$0.3$M_c$.

Of the remaining super-puffs, Kepler-51d and Kepler-87c cannot be explained by atmospheric masses $<$0.3$M_c$ assuming $T_{int}$ = 75 K. Smaller atmospheric mass fractions are possible if their cores contain more ices, such that they are more like water-worlds \citep{zeng2019}. Alternatively, they could possess higher $T_{int}$ due to a combination of tides and high masses. 

Kepler-11e, Kepler-223d, Kepler-177c, and Kepler-47c do not require hazes to simultaneously explain their masses, radii, and age, provided that their $T_{int}$ is closer to 75 K; including hazes changes their inferred atmospheric masses by $\sim$10-30\%. Non-super-puff sub-Neptunes experience reductions in inferred atmospheric mass by a few tens of \% as well if high altitude hazes were included but uncertainties in $T_{int}$ prevents more rigorous determinations.

Our mass-radius diagrams allow us to predict where more super-puffs may be found. Young planets with high $T_{int}$ are more likely to be super-puffs, since the high internal entropy promotes an extended atmosphere, while moderate outflow winds leads to the formation of transition hazes. If we focus specifically on planets with a few \% atmospheric mass fraction, which are more plentiful than objects with higher gas mass fractions \citep{owen2017}, then most super-puffs with $T_{eq}$ $\sim$ 700 K should have total mass between 2M$_{\Earth}$ and 4M$_{\Earth}$, with a peak in radius of $\sim$8R$_{\Earth}$ at 3M$_{\Earth}$, while for cooler objects the mass ranges should move to lower values due to the smaller scale heights and longer atmospheric lifetimes ($<$3M$_{\Earth}$ for $\sim$500 K objects, with a peak radius of $\sim$7R$_{\Earth}$ at 2M$_{\Earth}$, similar to Kepler-51b; $<$2.5M$_{\Earth}$ for $\sim$300 K objects, with a peak radius $>$6R$_{\Earth}$ at $\leq$1.5M$_{\Earth}$). Objects with higher core masses can retain higher gas masses and reach much larger radii, though they may be affected by stellar winds \citep{wang2018} and roche-lobe overflow \citep{rogers2011}, which we do not consider here. 

Table \ref{table:youngplanets} lists several planets that are intriguing candidates for super-puffs. They are all young planets with $T_{eq}$ $<$ 1000 K, and thus capable of hosting high altitude hazes that could be responsible for their large sizes. They have relatively weak mass constraints due to the difficulty of radial velocity mass measurements for young stars \citep[e.g.][]{crockett2012}, and thus could possess very low masses that hinder future mass determinations. If they are super-puffs, then they should have nearly-flat or sloped transmission spectra in the optical and near-IR wavelengths, with significantly smaller transit radii at mid-IR wavelengths (see ${\S}$\ref{sec:midir}).

\begin{figure}[hbt!]
\centering
\includegraphics[width=0.45 \textwidth]{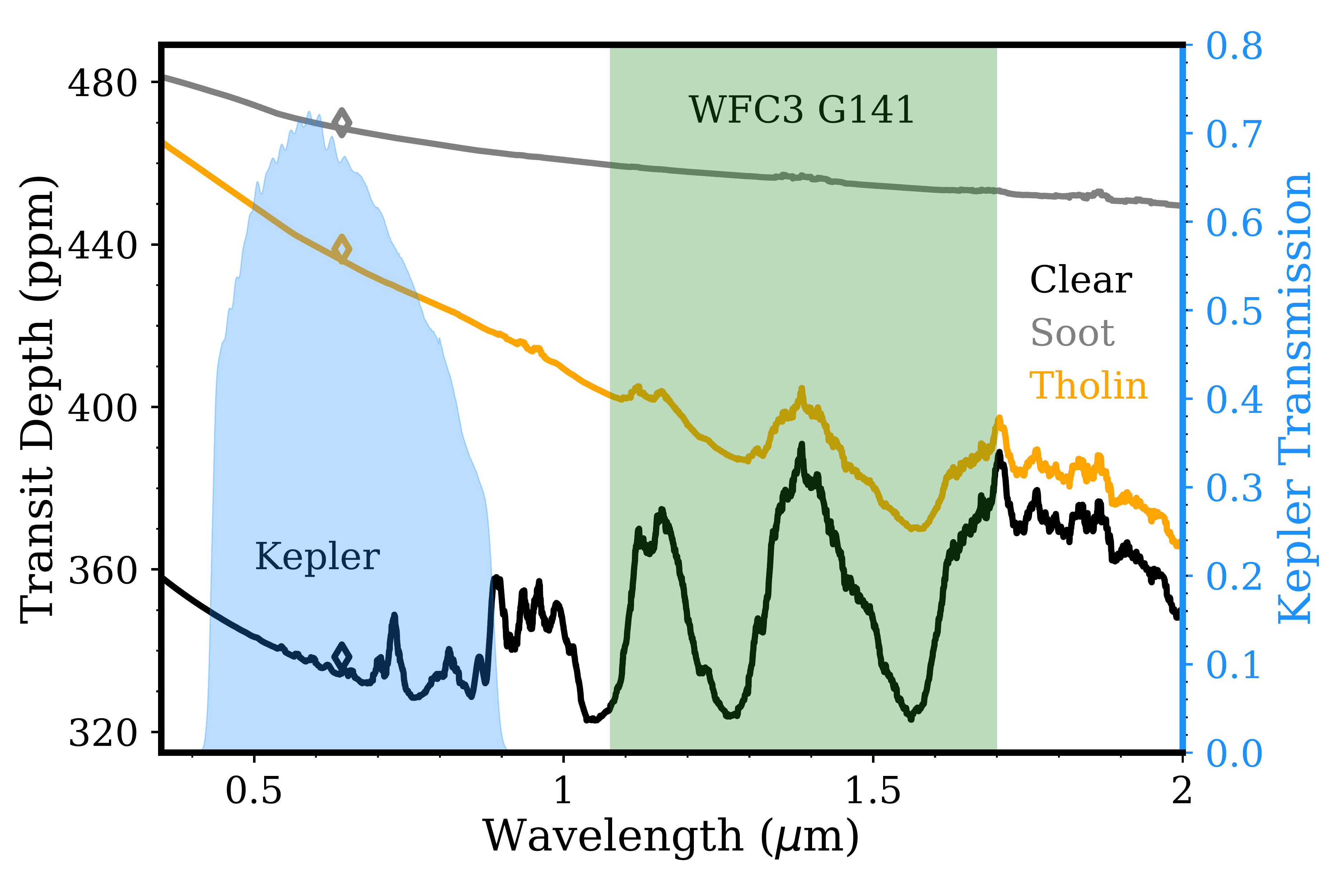}
\caption{Clear (black) and hazy (tholin: orange; soot: gray) model transmission spectra for a planet with $T_{eq}$ = 500 K, $T_{int}$ = 75 K, $M_c$ = 3M$_{\Earth}$, and $M_a$ = 0.01$M_c$. The \textit{Kepler} transmission function and HST WFC3 G141 wavelength range are indicated in the blue and green shaded regions, respectively. The transmission-weighted transit depth in the \textit{Kepler} band for the different transmission spectra are marked in diamonds.  }
\label{fig:kepler}
\end{figure}

\begin{figure*}[hbt!]
\centering
\includegraphics[width=0.7 \textwidth]{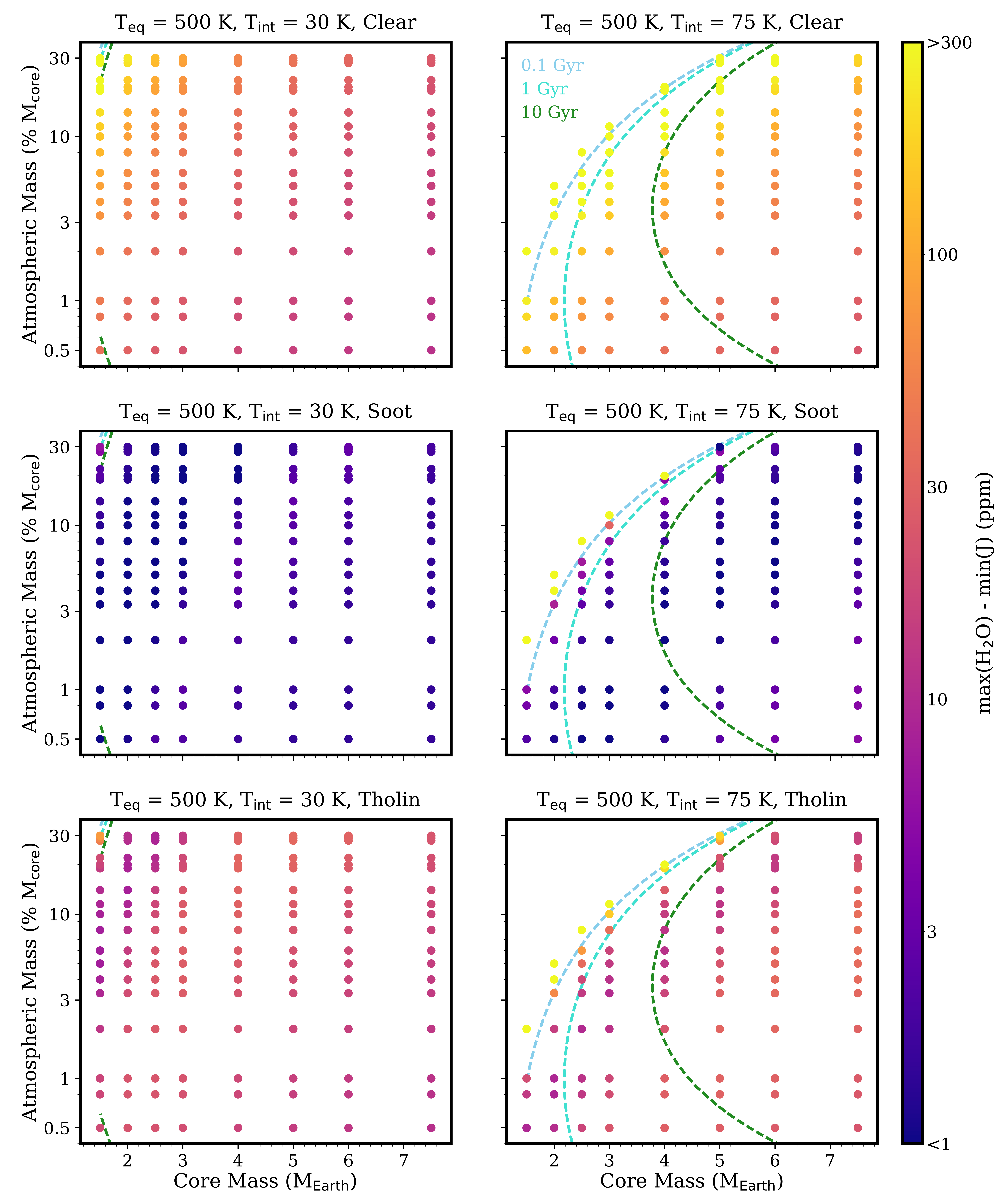}
\caption{The amplitude of the 1.4 $\mu$m water feature in transmission for clear (top) and hazy (soot: middle; tholin: bottom) objects with $T_{eq}$ = 500 K. Contours of atmospheric lifetimes are shown for 0.1 Gyr (light blue), 1 Gyr (cyan), and 10 Gyr (green). Results for other $T_{eq}$ cases can be found in Appendix \ref{sec:apph2oamp}.}
\label{fig:teq500h2oamp}
\end{figure*}

\subsection{Near-IR Transmission Spectroscopy}\label{sec:transmission}

We evaluate the effect of hazes on the near-IR transmission spectra of low mass planets, in particular the G141 band of \textit{Hubble Space Telescope's} Wide Field Camera 3 (Figure \ref{fig:kepler}), by computing model spectra at the resolution of the G141 grism and calculating the difference in transit depth between the maximum in the 1.4 $\mu$m water band (1.36-1.44 $\mu$m) and the minimum in the adjacent J band (1.22-1.36 $\mu$m). 

Our results show that absorbing (soot) transition hazes reduce the water feature amplitude to $<$1 ppm, consistent with the observations of \citet{libbyroberts2019}, while absorbing static hazes allow for water feature amplitudes between 3 and 10 ppm (Figure \ref{fig:teq500h2oamp}). Scattering (tholin) hazes show the same variations in water feature amplitude, but the amplitude itself is larger: $\sim$20 ppm for transition hazes and $\sim$50 ppm for static hazes. These results stem from two main factors: (1) the reduction in the pressures probed in transit due to haze opacity (Figure \ref{fig:teq500pressure}), and (2) the depletion of molecular absorbers above the haze due to photochemistry (Figure \ref{fig:tprof}). Both hazy cases contrast with the clear case, where the water feature amplitude increases monotonically with increasing $M_a$ and decreasing $M_c$, as expected from increasing scale heights due to decreasing gravity, reaching values $>$300 ppm in regions of the parameter space where transition hazes form. Short-lived planets featuring outflow hazes also exhibit large water feature amplitudes due to the low haze opacity and large scale heights. 

Our results suggest that sub-Neptunes with $T_{eq}$ $\sim$ 400-800 K are uniquely bad targets for identification of molecular features in optical and near-IR transmission spectra due to the presence of high altitude, opaque hazes, with younger planets (high $T_{int}$) having the smallest near-infrared water feature amplitudes due to transition hazes. In particular, if super-puffs' large radii are due to such hazes, then they would almost certainly have featureless transmission spectra in the near-IR. Warmer planets may avoid this issue due to the dominance of CO as the primary carbon reservoir curtailing haze production; this would be consistent with recent observations showing an increase in the 1.4 $\mu$m water feature amplitude of exo-Neptunes with increasing temperatures for $T_{eq}$ $>$ 600 K \citep{stevenson2016,fu2017,crossfield2017}. However, laboratory experiments have shown that CO can also act as haze parent molecules \citep{horst2018b,he2018,fleury2019}. 

Cooler planets in our model grid ($T_{eq}$ = 300 K) with static hazes have larger water feature amplitudes than the corresponding warmer planets, with amplitudes as large as 20 ppm with absorbing hazes and 50 ppm with scattering hazes. However, this may be caused by our choice of $K_{zz}$ enhancing molecular absorbers at low pressures (see ${\S}$\ref{sec:hazeprod}).

\subsection{The Case of Kepler-51b}\label{sec:kepler51b}

\begin{figure*}[hbt!]
\centering
\includegraphics[width=0.8 \textwidth]{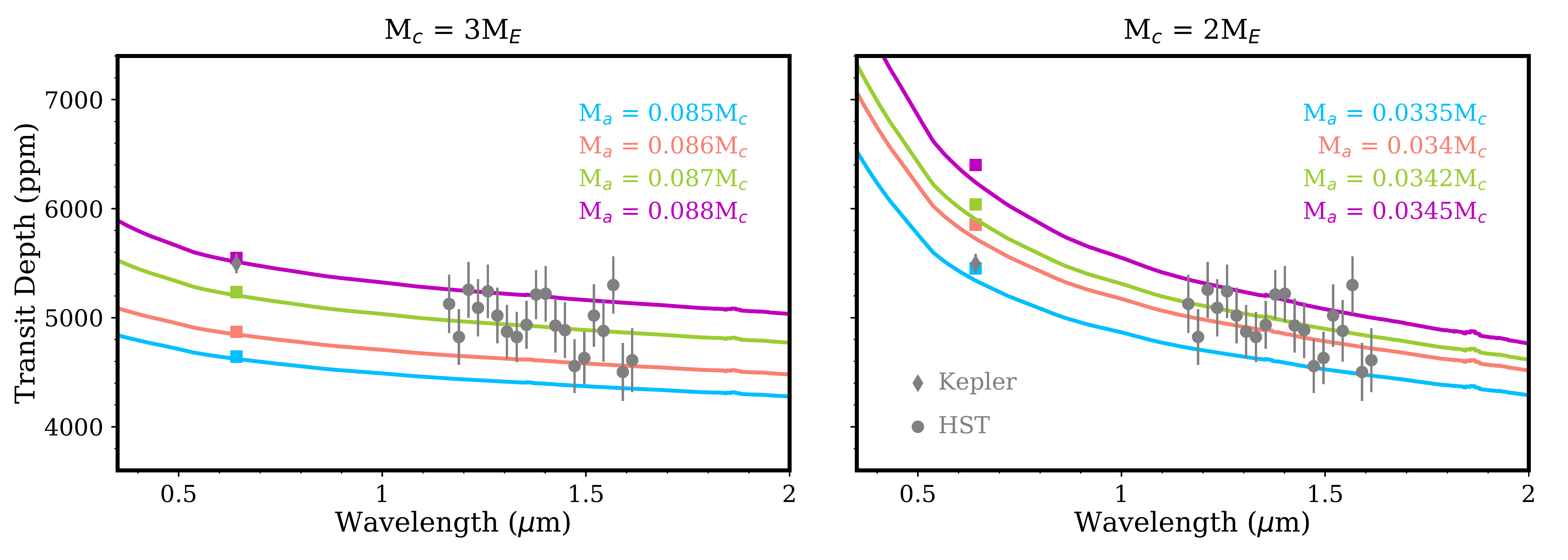}
\caption{Model transmission spectra of Kepler-51b assuming core masses of 3 (left) and 2 (right) Earth masses with various atmospheric mass fractions. The \textit{Kepler} transit depth \citep{masuda2014} (diamond) and \textit{HST} transmission spectrum \citep{libbyroberts2019} (circles) are shown in comparison. The transmission-weighted transit depth in the \textit{Kepler} band for the different transmission spectra are marked in squares. }
\label{fig:kepler51b}
\end{figure*}

Here we apply our modeling framework to Kepler-51b, one of the least dense super-puffs and one of the only ones with a near-IR transmission spectrum \citep{libbyroberts2019}. The gas mass fraction we derive from the available data is highly sensitive to the exact mass of the planet: for masses similar to that derived by \citet[Table \ref{table:superpuffs}]{libbyroberts2019}, we find $\overline{M}_a$ $\sim$ 9\% and an atmospheric lifetime of $\sim$0.5 Gyr, while for the lower mass (2.1 $^{+ 1.5 }_{ -0.8 }$ Earth masses) derived by \citet{masuda2014}, we find $\overline{M}_a$ $\sim$ 3\% and an atmospheric lifetime of $\sim$0.1 Gyr (Figure \ref{fig:kepler51b}). Both cases possess atmospheric lifetimes similar to the stellar age. We note, however, that the large uncertainties in both mass measurements mean that they are within 1$\sigma$ of each other. In addition, we caution that the apparent tightness of the gas mass fraction constraints provided by our modeling is misleading, as variations in the (unknown) atmospheric metallicity, haze optical properties, and other model parameters could change these values. The disagreement between models that best fit the near-IR spectrum and the optical \textit{Kepler} transit depth is expected given the high stellar activity of Kepler-51 \citep{libbyroberts2019}. Our gas mass fraction estimates are significantly lower than those of \citet{lopez2014} and \citet{libbyroberts2019}, who had $\overline{M}_a$ $>$ 10\%, though both of our works predict similar current atmospheric loss rates of 10$^{10}$-10$^{11}$ g s$^{-1}$. 

\section{Discussion}\label{sec:discussion}

\subsection{Implications for the Radius Evolution of Warm Sub-Neptunes}

Current formation and evolution theories of sub-Neptunes suggest that planets with rocky cores surrounded by gas envelopes with masses of a few \% of the core mass are one of the most common types of planets in the Galaxy \citep{fulton2017,owen2017,lee2016}. Super-puffs seemingly stand away from these objects by possessing gas envelopes with masses $>$10\% of the core mass, complicating their inferred formation process \citep{lee2016}. Our work does not change this notion for most super-puffs, as they all appear to require more than a few \% gas fraction even when high altitude hazes are included (though see ${\S}$\ref{sec:metallicity} for caveats). The only possible exception is Kepler-51b, the radius of which can be explained by $\leq$10\% gas fraction when hazes are taken into account. Kepler-223e may also fit this category, though its $T_{int}$ could be lower than what we have assumed.

The similarity of Kepler-51b's mass and derived gas fraction to those of the large radii population of sub-Neptunes suggests the intriguing possibility that (1) Kepler-51b will become part of this population after it has lost enough atmosphere to prolong its atmospheric lifetime past a few Gyr and cooled internally to a lower $T_{int}$, and that (2) lower mass ($<$4M$_{\Earth}$) members of this population may have been super-puffs earlier in their lives \citep{libbyroberts2019}. While the idea that planets lose atmosphere and contract over time as they cool is not new \citep[e.g.][]{lopez2014}, the inclusion of high altitude hazes means this radius evolution is more extreme: planets with masses $<$4M$_{\Earth}$ with only a few \% in gas mass could have evolved from Jupiter-size to near-Earth size over their first Gyr of life, rather than only $\sim$halving their radius.

One caveat of our results is the degeneracy between atmospheric mass fraction and $T_{int}$. Our model does not treat the thermal evolution of the interiors of low mass planets, and therefore we cannot say with certainty whether the $T_{int}$'s we have chosen reflect the actual intrinsic luminosity of the planets we are modeling. Therefore, one way to reduce our computed gas mass fractions of super-puffs aside from Kepler-51b to values more similar to the large radii population of sub-Neptunes is if we underestimated their $T_{int}$ values.


\subsection{Sensitivity Tests}

In deriving the radius enhancement caused by high altitude photochemical hazes, we assumed fixed values for the eddy diffusion coefficient ($K_{zz}$; 10$^8$ cm$^2$ s$^{-1}$), the atmospheric metallicity (solar), and the haze production efficiency ($\epsilon_h$ = 0.1). In reality all of these quantities can vary by at least an order of magnitude. Here we evaluate the sensitivity of our results to changes in these parameters. 

\subsubsection{$K_{zz}$}

\begin{figure}[hbt!]
\centering
\includegraphics[width=0.45 \textwidth]{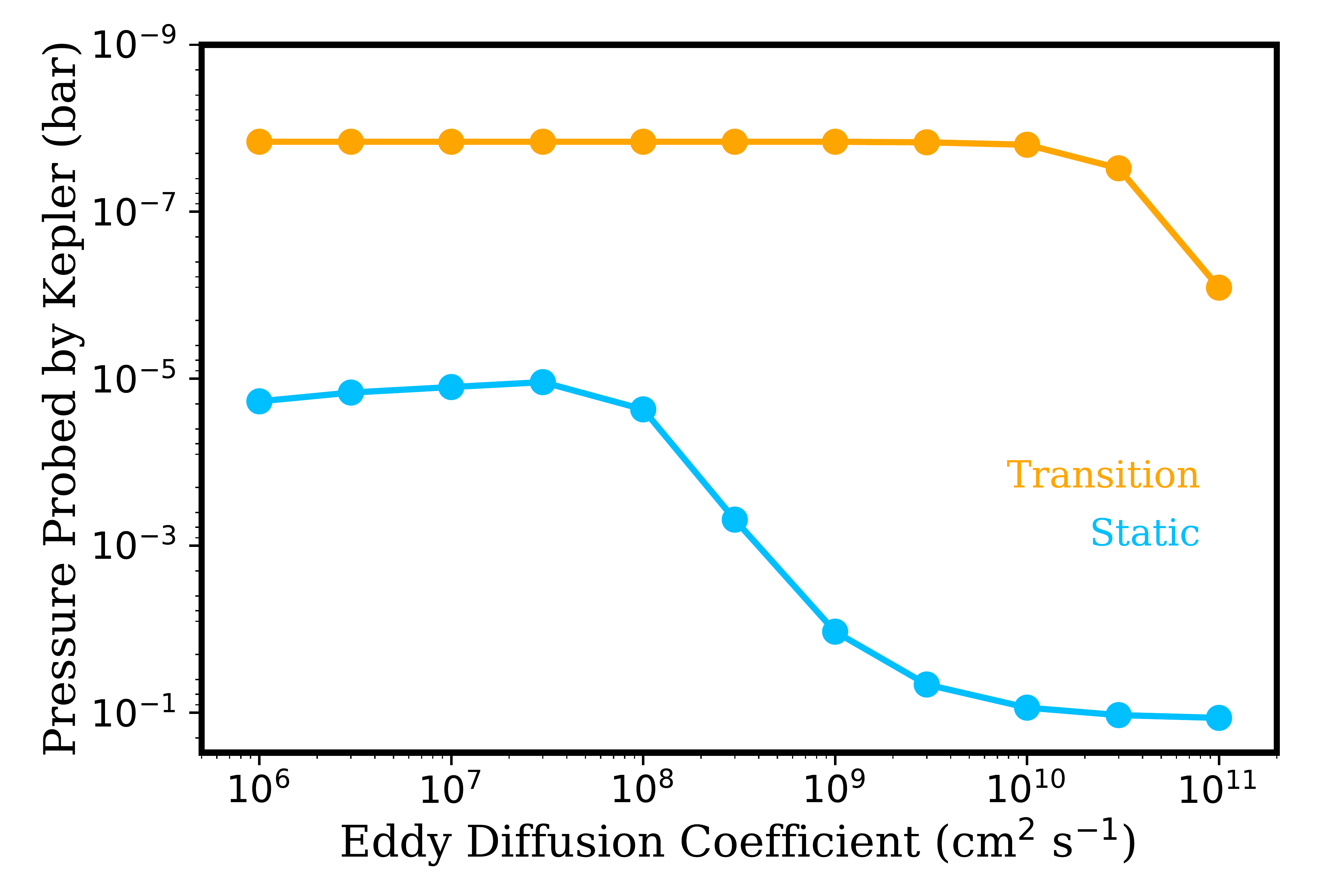}
\caption{Sensitivity of pressure probed in the \textit{Kepler} bandpass to variations in $K_{zz}$ for the same static (blue) and transition (orange) hazes as in Figure \ref{fig:hazedist}. }
\label{fig:kzzvar}
\end{figure}

As briefly discussed in ${\S}$\ref{sec:composition}, the eddy diffusion coefficient is a parameterization of large scale transport in an atmosphere. Values and profiles of $K_{zz}$ used in previous exoplanet modeling works range across several orders of magnitude \citep[e.g.][]{moses2011}, while measured $K_{zz}$'s of solar system objects also vary by similar amounts \citep{zhang2018a}. $K_{zz}$ may also not be appropriate for use in modeling photochemical hazes, particularly for tidally locked planets, due to upwelling plumes concentrating hazes at high altitudes on the dayside rather than diluting them through mixing \citep{zhang2018b}. While the 700 and 500 K planet cases may be tidally locked given a sun-like host star, the 300 K planets may be sufficiently far away to retain its primordial rotation period \citep{barnes2017}. Upwelling on the dayside of tidally locked planets would act to enhance haze optical depth by keeping aloft larger particles, similar to the effect of an outflow wind. 

Transition hazes, and thus our super-puff results, are insensitive to $K_{zz}$ variations for $K_{zz}$ $<$ 10$^{10}$ cm$^2$ s$^{-1}$ (Figure \ref{fig:kzzvar}). This is because the resupply of methane to the haze production region is dominated by the outflow wind. This changes for $K_{zz}$ $>$ 10$^{10}$ cm$^2$ s$^{-1}$, when transport in the haze production region becomes dominated by mixing, quenching haze particle growth by coagulation and resulting in smaller particles and a reduction in haze opacity. In other words, haze particles are mixed into the deep atmospheres before they can grow. Static hazes behave similarly to transition hazes, but for different reasons. While mixing resupplies the upper atmosphere with methane, the haze production region does not move appreciably as it is set by the Lyman-$\alpha$ opacity of methane and water, which increase rapidly with increasing depth in the atmosphere. The reduction in haze opacity also occurs at lower $K_{zz}$ ($\sim$10$^{8}$ cm$^2$ s$^{-1}$) due to the smaller scale heights of these atmospheres reducing the mixing timescale. 


\subsubsection{Metallicity}\label{sec:metallicity}

\begin{figure}[hbt!]
\centering
\includegraphics[width=0.45 \textwidth]{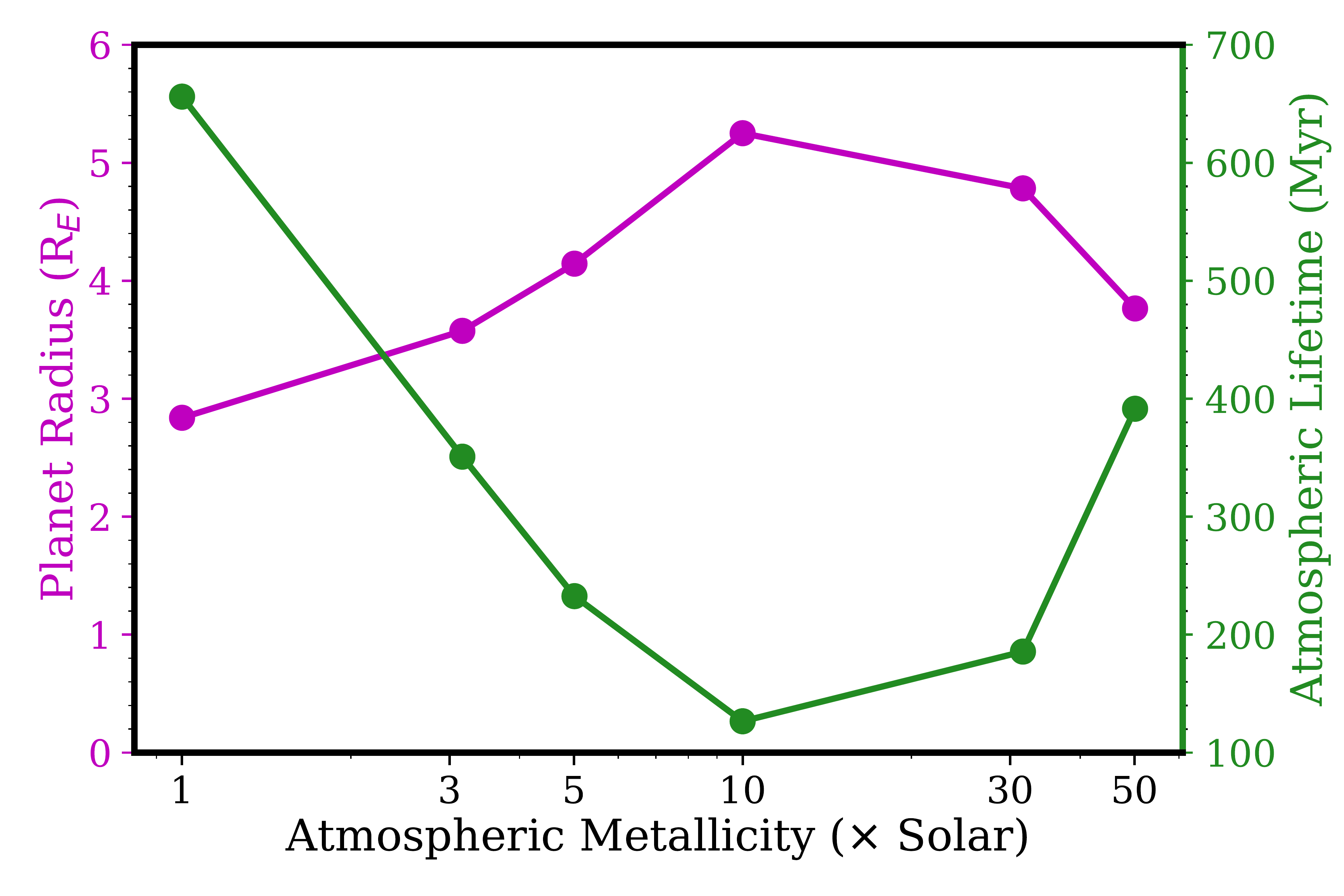}
\caption{The planet radius in the \textit{Kepler} bandpass for a planet with $T_{eq}$ = 500 K, $T_{int}$ = 75 K, $M_a$ = 0.01$M_c$, and $M_c$ = 2M$_{\Earth}$ as a function of metallicity (magenta) and the corresponding atmospheric lifetimes (green).  }
\label{fig:metvar}
\end{figure}

\begin{figure*}[hbt!]
\centering
\includegraphics[width=0.8 \textwidth]{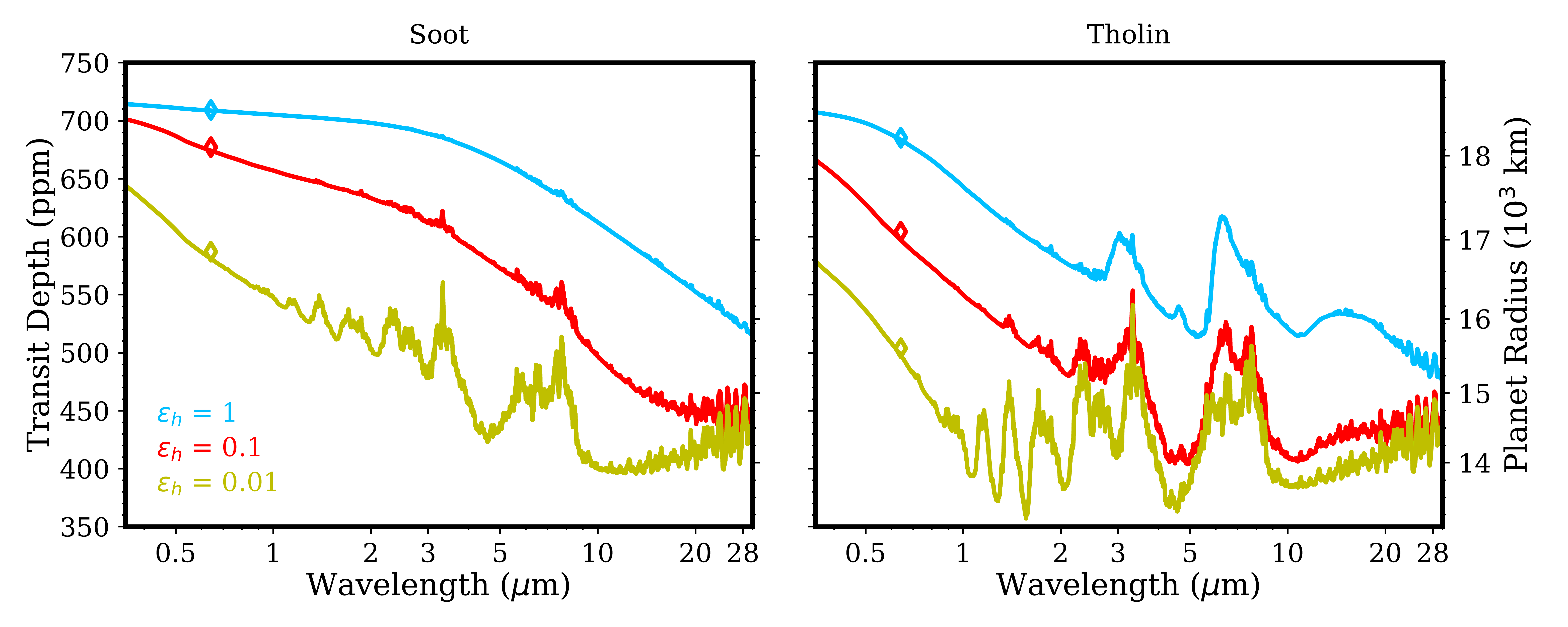}
\caption{Transmission spectra for a planet with $T_{eq}$ = 500 K, $T_{int}$ = 75 K, $M_a$ = 0.01$M_c$, and $M_c$ = 2M$_{\Earth}$ with haze production efficiencies of 0.01 (yellow), 0.1 (red), and 1 (blue) for soot (left) and tholin (right) hazes. The Kepler bandpass-averaged planet radii are shown by diamonds.}
\label{fig:effvar}
\end{figure*}

There have been relatively few measurements of the metallicity of sub-Neptunes and Neptune-mass exoplanets, with the available observations showing great diversity. Several objects show flat transmission spectra indicative of high altitude aerosols, metallicities greater than 1000 $\times$ solar, and/or a lack of an atmosphere altogether \citep{kreidberg2014,knutson2014a,knutson2014b,dewit2018}. In contrast, other objects show metallicities below 100 $\times$ solar, and some even approaching 1 $\times$ solar \citep{wakeford2017b,benneke2019}. Still others exhibit spectral features indicating metallicities between 1 and 1000 $\times$ solar \citep[e.g.][]{fraine2014}. 

We use the $>$1 $\times$ solar metallicity opacity tables of \citet{freedman2014} and the corresponding (higher) atmospheric mean molecular weights to explore the effect of increasing atmospheric metallicities on our results (Figure \ref{fig:metvar}). Increasing metallicities increases the opacity of the atmosphere, and thus the radiative convective boundary is reached at lower pressures (Eq. \ref{eq:rdgrad}), leading to a warmer interior for a fixed temperature at the RCB. For metallicities $\leq$10 $\times$ solar, this leads to a larger planet, and thus a lower atmospheric mass is needed to achieve a given planet radius compared to the solar metallicity case. For example, for a planet similar to Kepler-51b but with an atmosphere that is only 1\% the mass of the core, increasing metallicity from solar to 10 $\times$ solar nearly doubles the planet radius. This causes a corresponding decrease in the atmospheric lifetime due to increased atmospheric density at the exobase. At higher metallicities ($>$10 $\times$ solar), the planet radius is reduced due to decreasing scale heights. Therefore, super-puffs should have low metallicity ($\leq$50 $\times$ solar), or else they must possess a much more massive gas envelope and/or a higher $T_{int}$. Clear planets should exhibit a similar trend as hazy planets, though with all planet radii shifted to smaller values. 

\subsubsection{Haze Production Efficiency}\label{sec:effvar}

The haze production efficiency is highly uncertain, being a detailed function of complex photochemical reactions. Previous works that converted photochemical modeling results to haze production rates (see ${\S}$\ref{sec:hazeprod}) considered haze efficiencies of 0.01-0.1. Direct comparisons between our work and theirs are difficult, however, due to our reliance on methane exclusively, while other works also take into account nitrogen species like HCN, thereby increasing the mass of haze precursors. As such, $\epsilon_h$ $>$ 0.1 when applied to only methane photolysis is certainly possible. By increasing $\epsilon_h$ from 0.01 to 1, we find increasing enhancement in planet radius in the \textit{Kepler} band and flatter near-IR transmission spectra (Figure \ref{fig:effvar}). The magnitude of the change in planet radius is $\sim$15-40\% depending on the haze optical properties and wavelengths of observation.

\subsection{Haze Effects at Longer Wavelengths}\label{sec:midir}

\begin{figure}[hbt!]
\centering
\includegraphics[width=0.45 \textwidth]{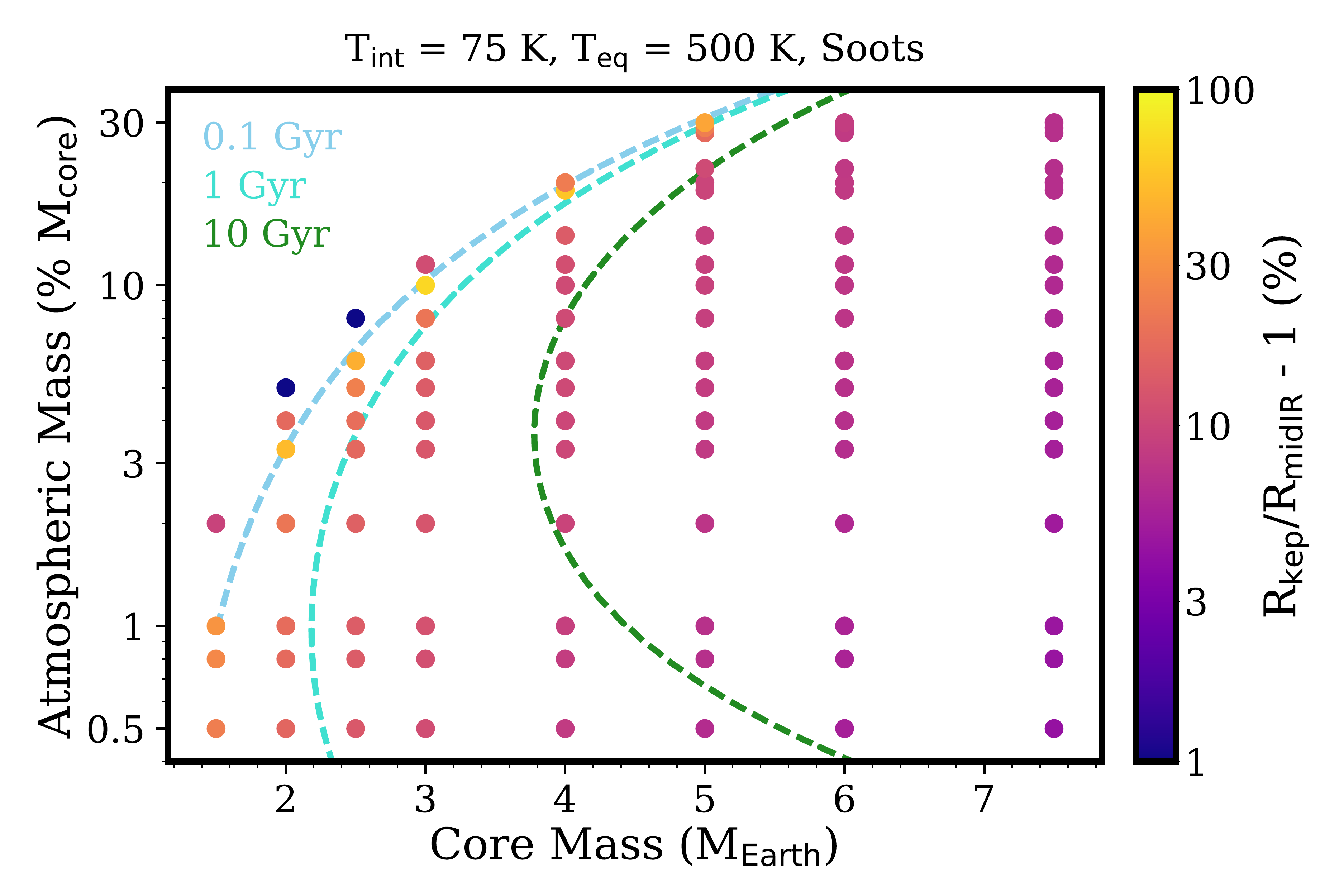}
\caption{The increase in the radius of the planet in the optical versus the mid-infrared wavelengths (10-12 $\mu$m) due to soot hazes for planets with $T_{eq}$ = 500 K and $T_{int}$ = 75 K. Contours of atmospheric lifetimes are shown for 0.1 Gyr (light blue), 1 Gyr (cyan), and 10 Gyr (green). }
\label{fig:midir}
\end{figure}

\begin{figure}[hbt!]
\centering
\includegraphics[width=0.45 \textwidth]{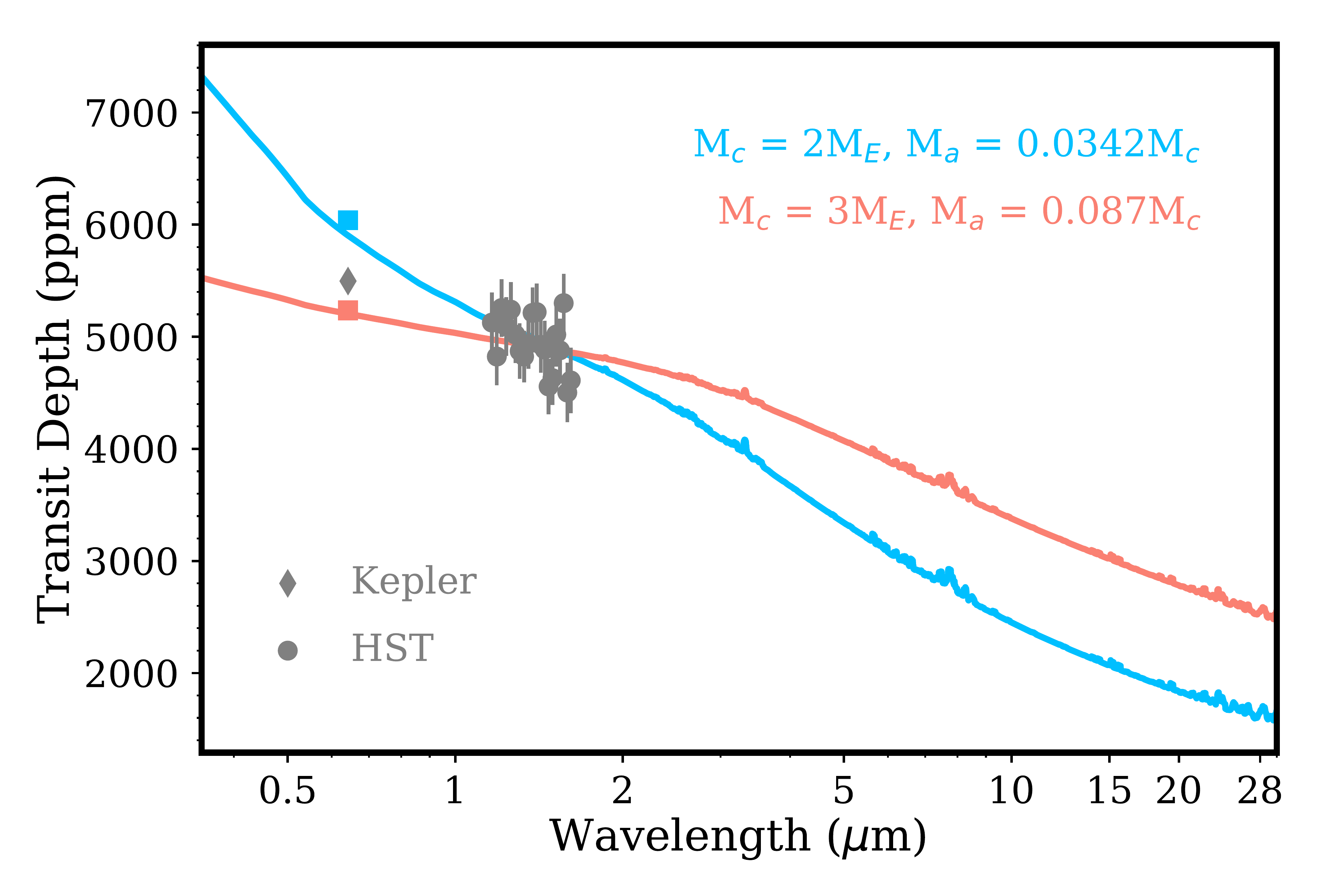}
\caption{Model transmission spectra of Kepler-51b assuming core masses of 3 (red) and 2 (blue) Earth masses with  atmospheric mass fractions that best fit the \textit{HST} data (circles). The \textit{Kepler} transit depth \citep{masuda2014} (diamond) is also shown in comparison. The transmission-weighted transit depth in the \textit{Kepler} band for the different transmission spectra are marked in squares.}
\label{fig:kep51bmidir}
\end{figure}

The James Webb Space Telescope (JWST) will be able to observe sub-Neptunes from 0.6 to 12 $\mu$m in transit using a suite of instruments \citep{greene2016}, and as such it is important to predict the effect of hazes at wavelengths $>$2 $\mu$m. As shown in Figure \ref{fig:effvar}, the decreasing opacity of spherical Mie particles with increasing wavelength could result in planet radii that are much larger when viewed in the optical versus the mid-IR. This effect increases in magnitude with increasing outflow wind speed, and could reach a factor of 2 in the transition haze region (Figure \ref{fig:midir}). This is due to the large atmospheric scale height and size sorting of aerosols by altitude. As a result, future observations of super-puffs using JWST must take into account the possibility that the planet is significantly smaller at wavelengths longer than that already observed by \textit{Kepler}. Figure \ref{fig:kep51bmidir} shows our prediction for the mid-IR transmission spectra of Kepler-51b using the two best fit models from Figure \ref{fig:kepler51b}, where the mid-IR transit depth is half that of the near-IR. The spectra are also nearly featureless due to the smoothly varying refractive indices of soot, though there is a significantly spectral slope and a $\sim$100 ppm methane feature at 7.6 $\mu$m. In contrast, tholin-like hazes may exhibit larger features (Figure \ref{fig:effvar}), including at 3 and 7 $\mu$m, which should be detectable by JWST, though the strength and location of the spectral features of actual exoplanet photochemical hazes are uncertain \citep{he2018}.

\subsection{Haze Radiative Feedback}

\begin{figure}[hbt!]
\centering
\includegraphics[width=0.45 \textwidth]{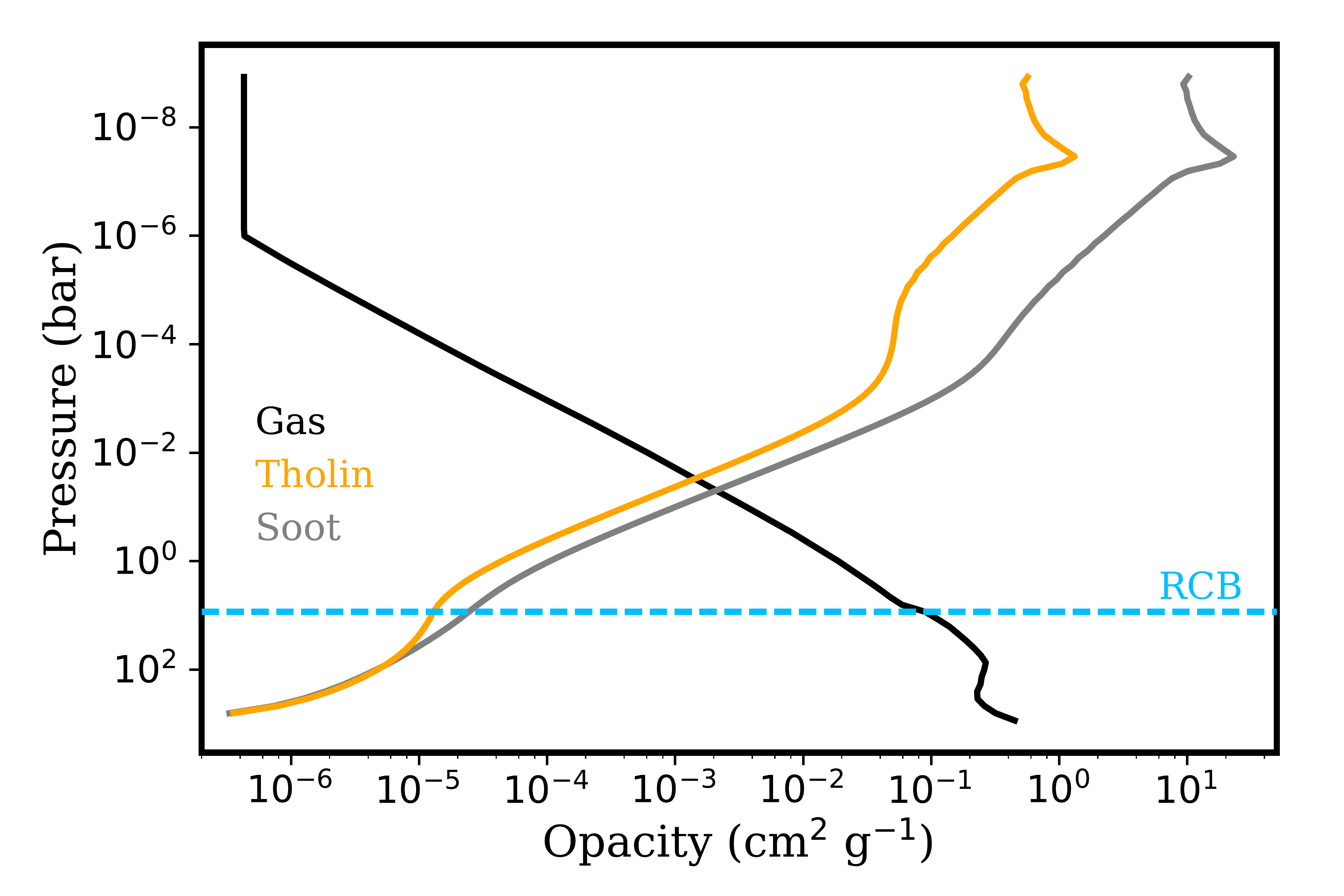}
\caption{Rosseland mean opacity for gases (black) and soot (gray) and tholin (orange) hazes for a planet with $T_{eq}$ = 500 K, $T_{int}$ = 75 K, $M_a$ = 0.033$M_c$, and $M_c$ = 2M$_{\Earth}$. The constant gas opacity at pressures $<$1 $\mu$bar is due to extrapolation of the opacity table from \citet{freedman2014} and should not affect our main results, since the opacity is mostly important in determining the location of the radiative-convective boundary (blue dashed line), located at much higher pressures. }
\label{fig:opacity}
\end{figure}

The dominance of the haze in transmission spectra suggests that the haze opacity should be much larger than the gas opacity. We calculate the Rosseland mean opacity of soot and tholin hazes following the procedure of \citet{freedman2014} over the same wavelength range and compare them to the gas opacity, shown in Figure \ref{fig:opacity}. The haze opacity, regardless of composition, is $\sim$6 orders of magnitude larger than the gas opacity at pressures $<$1 mbar, and thus hazes are the main controls of the upper atmosphere temperature structure and radiation field. It is uncertain, however, whether hazes will heat or cool the atmosphere; previous works that consider haze feedback have shown that haze heating could lead to a temperature inversion that results in observable emission features \citep[e.g.][]{morley2015}, but haze cooling may also be important \citep{zhang2015,zhang2017}. As the upper atmosphere temperature dictates the rate of atmospheric loss and photochemical reactions, understanding the radiative effects of haze could be vital in constraining the atmospheric evolution and composition of warm sub-Neptunes. 

In contrast, as the haze opacity is much smaller than the gas opacity at pressures $>$1 bar, the location of the RCB, and thus the internal entropy of the planet, is unlikely to be affected. These conclusions are independent of whether the haze is static or transition, as the haze opacity is always much larger than the gas opacity at low pressures.




\subsection{Additional Photochemical Considerations}\label{sec:photocons}

Our treatment of methane photochemistry is highly simplified, relying only on Lyman-$\alpha$ photons and opacity from methane and water vapor. Despite this simplicity, however, the computed pressure levels where haze formation peaks are similar between our work and those relying on more sophisticated photochemical models \citep[$\sim$1 $\mu$bar, e.g.][]{kawashima2019a,lavvas2019} that  take into account the full stellar spectrum and the shielding effects of other molecules, including photodissocation products. However, what we do not capture is the more extended region of haze formation at higher pressures. This is due to our neglect of the photodissociation of methane and water vapor by lower energy photons, which penetrate deeper into the atmosphere and can have wavelengths as long as $\sim$250 nm \citep{adamkovics2014,vanharrevelt2006}. This suggests that including longer wavelength photons would result in higher rates of haze production throughout the atmosphere, further boosting the radii of haze-hosting low mass planets. 

Another complication to our treatment of photochemistry is the opacity of atomic and excited molecular hydrogen \citep{bethell2011}, which could be abundant in photoevaporating atmospheres \citep{wang2018}. These species would act to push the Lyman-$\alpha$ photosphere to lower pressures. The effect of this on our results depends on whether the enlarged Lyman-$\alpha$ photosphere lies above or below the pressure level of diffusive separation, and how it varies due to the outflow. If the Lyman-$\alpha$ photosphere remains in the well-mixed part of the atmosphere, then the haze layer would simply form at a lower pressure, further enlarging the planet; otherwise, haze formation may be reduced due to the decrease in haze parent molecule abundance. A photochemical model that takes into account a photoevaporative outflow is needed to investigate this in detail, which is beyond the scope of this study. 

\subsection{Additional Microphysical Considerations}

In our work we consider only spherical haze particles. However, \citet{adams2019} showed that haze particles composed of fluffy aggregates of small monomers can form in exoplanet atmospheres, and that they can significantly increase haze opacity and reduce the wavelength dependence of haze opacity compared to a case with the same haze production rate of spherical particles. However, the increase in opacity is dependent on the porosity of the aggregates, with higher porosity leading to decreased opacity. \citet{ohno2019} showed that compression forces in exoplanet atmospheres are unlikely to strongly impact aggregates, and that they will maintain a fractal dimension of $\sim$2, rather than 2.4 as assumed in \citet{adams2019}. This leads to a wavelength-dependence of the opacity more similar to that of the individual monomers, which are small enough to create spectral slopes \citep{lavvas2019}. Given the uncertainties regarding how aggregates form and evolve in exoplanet atmospheres, it is difficult to deduce how they would affect the observed planet radius. 

Another possible control on the haze distribution is the impact of condensation. Falling photochemical haze particles can act as condensation nuclei for water clouds on Earth \citep{boucher1995} and hydrocarbon clouds on Titan \citep{lavvas2011}. On sub-Neptunes, the possible condensates are chlorine salts like KCl and sulfides like ZnS and Na$_2$S \citep{morley2012}. The pressure levels where these clouds form depends on the exact temperature-pressure profile, while the nucleation rates of clouds on haze particles depends on their material properties. The effect of nucleation would be to increase the aerosol opacity of the atmosphere by conversion of condensate vapor to solids/liquids, though this will mostly impact higher pressures near the condensate cloud bases, and thus should not affect our results.

\section{Summary and Conclusions}\label{sec:conclusions}

The nature of super-puffs is difficult to explain due to an inability to simultaneously reconcile their observed radii, masses, inferred gas mass fractions, and atmospheric lifetimes given clear atmospheres. We have shown using a suite of atmospheric and aerosol microphysical models that high-altitude photochemical hazes, such as that found on Titan and Pluto, could provide a natural solution. Hazes can explain not only the inflated radii of some super puffs in the optical, but also the flat transmission spectra of several super-puffs seen in the near-infrared. Furthermore, we have extended our modeling framework to warm and temperate low-mass planets in general, which allows us to conclude the following:


\begin{itemize}
  \item Haze opacity is enhanced by the outflow wind due to (1) rapid replenishment of methane lost to photolysis in the upper atmosphere, pushing the haze formation region to higher altitudes, and (2) reduced sedimentation velocity of haze particles leading to larger particles at low pressures. Maximum haze opacity (``transition hazes'') is achieved when the wind speed is such that 10-100 nm particles have near-zero net velocity in the haze formation region, which implies atmospheric lifetimes of 0.1-1 Gyr. Slower winds (longer lifetimes) lead to haze formation deeper in the atmosphere and smaller particles, while faster winds suppress haze particle growth. 
  \item The inclusion of high altitude hazes decreases the pressure probed in transmission, with young (0.1-1 Gyr), warm (T$_{eq}$ $\geq$ 500 K), and low mass ($M_c$ $<$ 4M$_{\Earth}$) objects experiencing the largest decrease due to hosting transition hazes, reaching $\sim$10-100 nbar. This directly translates to an increase in the observed transit radius with respect to the clear atmosphere case, which can reach several hundred \%, thus doubling or tripling the clear atmosphere radius. 
  \item Planets with masses $<$10M$_{\Earth}$, ages $\leq$1 Gyr, and $T_{eq}$ $\sim$ 400-800 K should possess sufficiently opaque hazes at low pressures to reduce the 1.4 $\mu$m water feature amplitude to $\leq$1 ppm for absorbing hazes and $\leq$30 ppm for scattering hazes. 
  \item Our results reconcile the observed masses, radii, and ages of several super-puffs, including Kepler-51b and c, by reducing the pressures probed in the \textit{Kepler} bandpass to several nbar rather than $\sim$100 mbar. This has the effect of increasing their atmospheric lifetimes to $>$0.1 Gyr, consistent with their host star's inferred age, and reducing the gas mass fraction of Kepler-51b to $\leq$10\%. A high altitude haze also explains the featureless transmission spectrum of Kepler-51b \citep{libbyroberts2019}. The effect on other super-puffs are less clear-cut due to uncertainties in $T_{int}$. 
  \item The low gas mass fraction inferred for a hazy Kepler-51b places it closer to objects on the large radius side of the sub-Neptune radius valley, suggesting that it may evolve to become part of that population after simultaneously losing some of its atmosphere and cooling internally. Conversely, objects on the large radius side of the radius valley that are cool enough to host hazes may have once been super-puffs in their early evolution. This suggests that the radius evolution of sub-Neptunes could span an order of magnitude even well after their birth, and that currently known young Jupiter-sized planets with weak mass constraints may be hidden super-puffs.
  \item Sensitivity tests show that our explanation for the nature of super-puffs is independent of the eddy diffusion coefficient for values $<$ 10$^{10}$ cm$^2$ s$^{-1}$, while variations of the haze production efficiency from 0.01 to 1 can alter the optical transit depth by tens of \%. 
  \item The planet radius and atmospheric lifetime for a given gas mass fraction exhibit non-linear dependencies on the atmospheric metallicity, as  metallicity affects both the opacity and the atmospheric scale height. Increasing atmospheric metallicity up to 10 $\times$ solar reduces the gas masses needed to account for a given planet radii due to higher temperatures in the convective envelope, while increasing metallicity further reverses this effect, as the atmospheric scale height decreases from the increased mean molecular weight. Thus, super-puffs should have relatively low metallicities ($<$50 $\times$ solar), or else much higher gas masses and/or intrinsic luminosity. 
  \item The decrease in haze opacity at longer wavelengths means that observations in the mid-infrared by JWST must account for a significantly smaller planet, sometimes by as much as a factor of 2. 
  \item Atmospheric opacity is dominated by hazes at pressures $<$1 mbar, suggesting that hazes control the upper atmosphere temperature structure. A temperature inversion may be possible if haze heating is significant. On the other hand, the atmospheric loss rate could be reduced if the haze cooling is significant.

\end{itemize}

Our work shows the significance of connecting atmospheric processes to planetary evolution: the haze opacities required to explain the large radii of some super-puffs are only possible due to the enhancing effects of winds associated with atmospheric loss. These objects are thus ephemeral, as they will evolve significantly over the next Gyr. Capturing this evolution, as well as clarifying the nature of super-puffs with high gas masses, require coupling the thermal evolution of these objects with the evolution of haze opacity, and evaluating the impact of high altitude hazes on atmospheric escape. 

Improved mass constraints and future observations by JWST and other observational platforms at longer wavelengths and other observational geometries (i.e. emission or reflection) will be vital for testing our hypothesis. The reduced opacities of the haze at longer wavelengths and in nadir geometries should reveal a much smaller planet and more gas spectral features, while reflected light measurements can help identify the nature of the haze \citep{morley2015,fortney2019}. Observations in the UV should shed light on the intensity of atmospheric escape, as well as the possible entrainment of haze particles in the escaping wind. Thus, super-puffs would greatly benefit from a comprehensive observing strategy, which will allow us to fully understand these unique worlds balanced on the edge of stability.  

\acknowledgments

We thank E. D. Lopez, J. E. Libby-Roberts, C. V. Morley, S. Ginzburg, D. P. Thorngren, J. E. Owen, and K. Ohno for enlightening discussions. We thank H. Zhang and W. Z. Gao for their loving support during the writing of this paper. The idea for this work was conceived at the 2018 Exoplanet Summer Program administered by the Outer Worlds Laboratory at the University of California, Santa Cruz, funded by the Heising-Simons Foundation. P. Gao acknowledges support from the 51 Pegasi b Fellowship, also funded by the Heising-Simons Foundation. X. Zhang is supported by NASA Solar System Workings Grant 80NSSC19K0791.



\vspace{5mm}


\bibliography{references}
\bibliographystyle{aasjournal}

\appendix

\section{Optical Transit Pressures and Radii at 700 and 300 K}\label{sec:appprerad}

\begin{figure*}[hbt!]
\centering
\includegraphics[width=0.7 \textwidth]{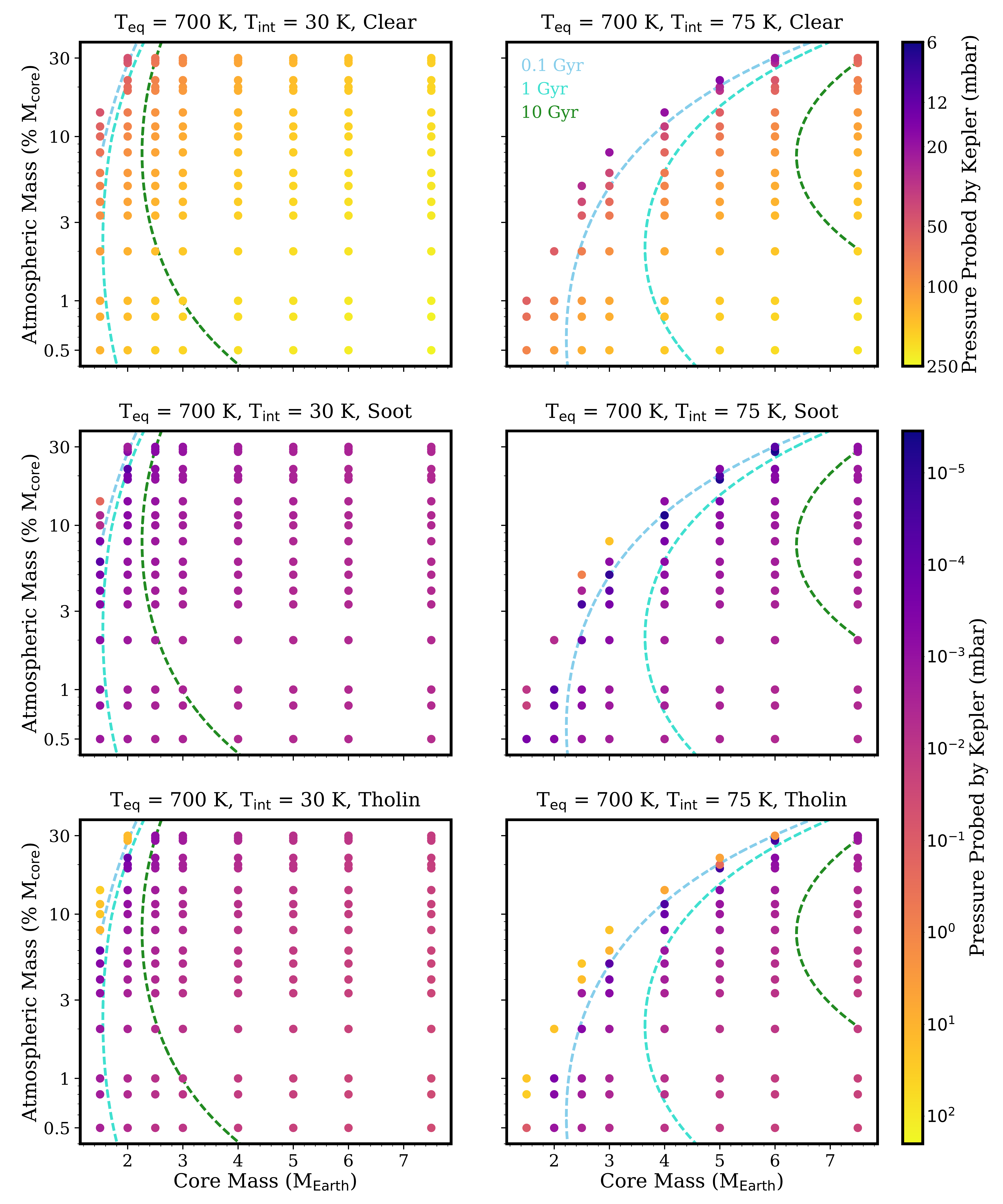}
\caption{Same as Figure \ref{fig:teq500pressure}, but for 700 K cases.}
\label{fig:teq700pressure}
\end{figure*}

\begin{figure*}[hbt!]
\centering
\includegraphics[width=0.7 \textwidth]{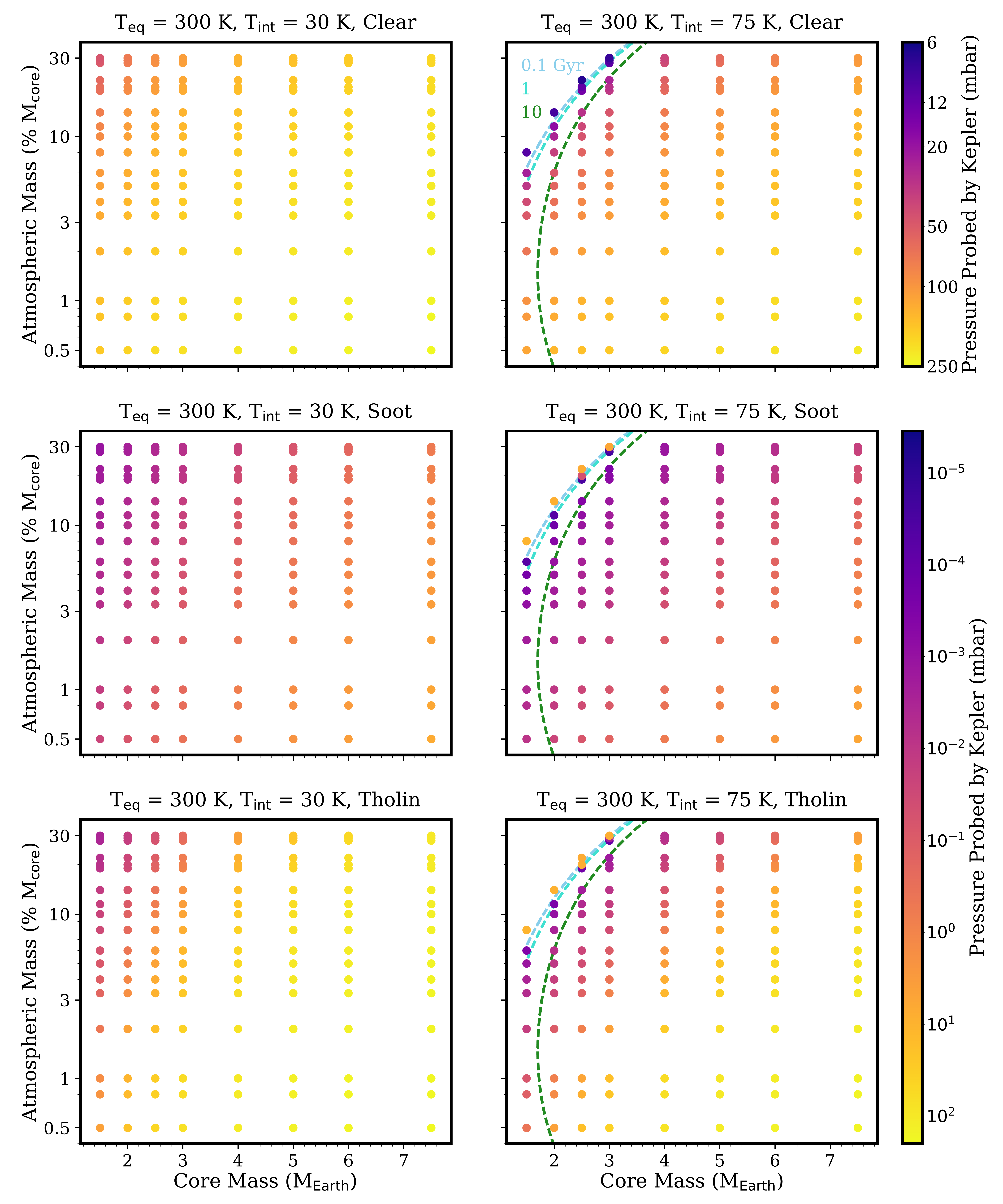}
\caption{Same as Figure \ref{fig:teq500pressure}, but for 300 K cases.}
\label{fig:teq300pressure}
\end{figure*}

\begin{figure*}[hbt!]
\centering
\includegraphics[width=0.7 \textwidth]{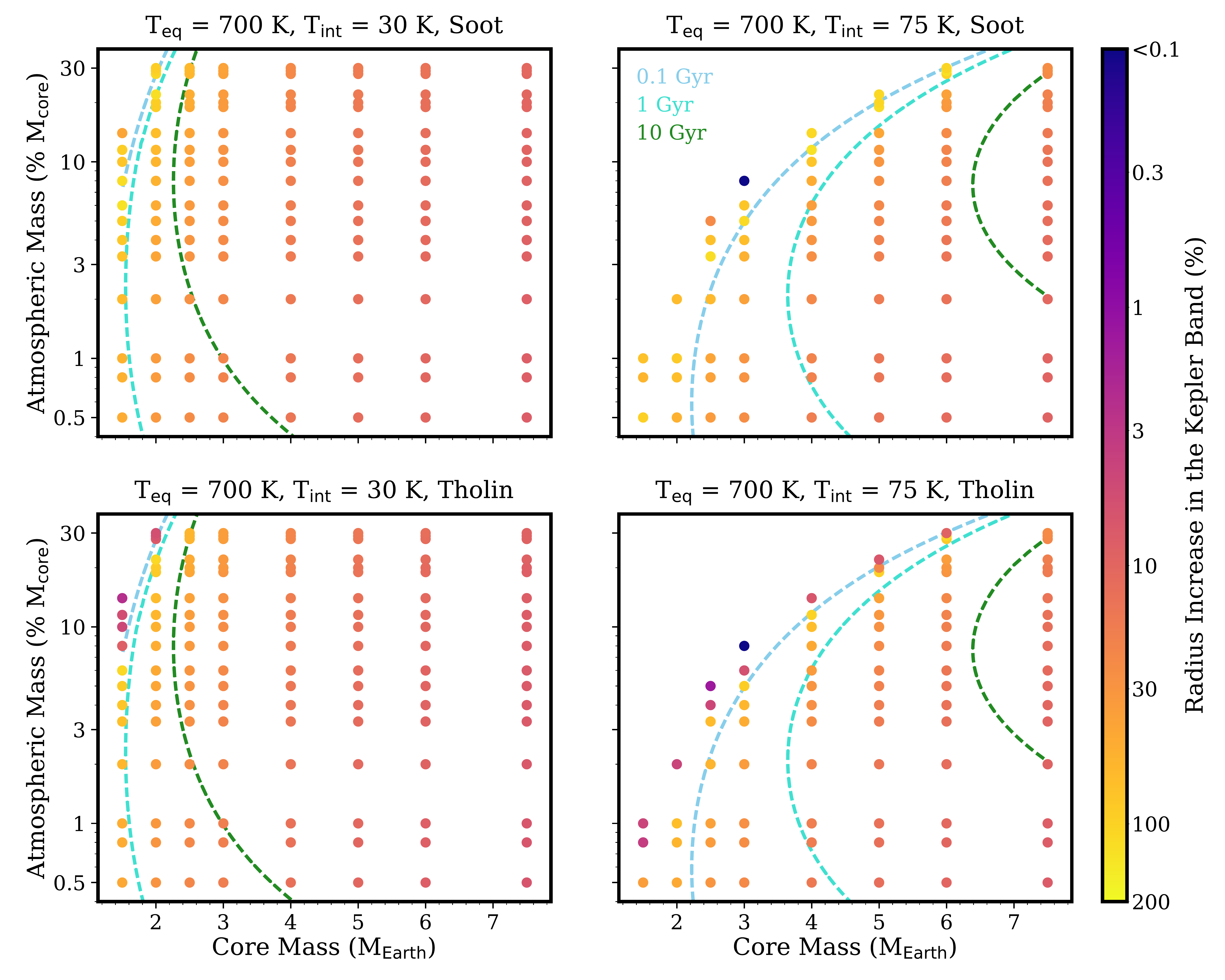}
\caption{Same as Figure \ref{fig:teq500raddif}, but for 700 K cases.}
\label{fig:teq700raddif}
\end{figure*}

\begin{figure*}[hbt!]
\centering
\includegraphics[width=0.7 \textwidth]{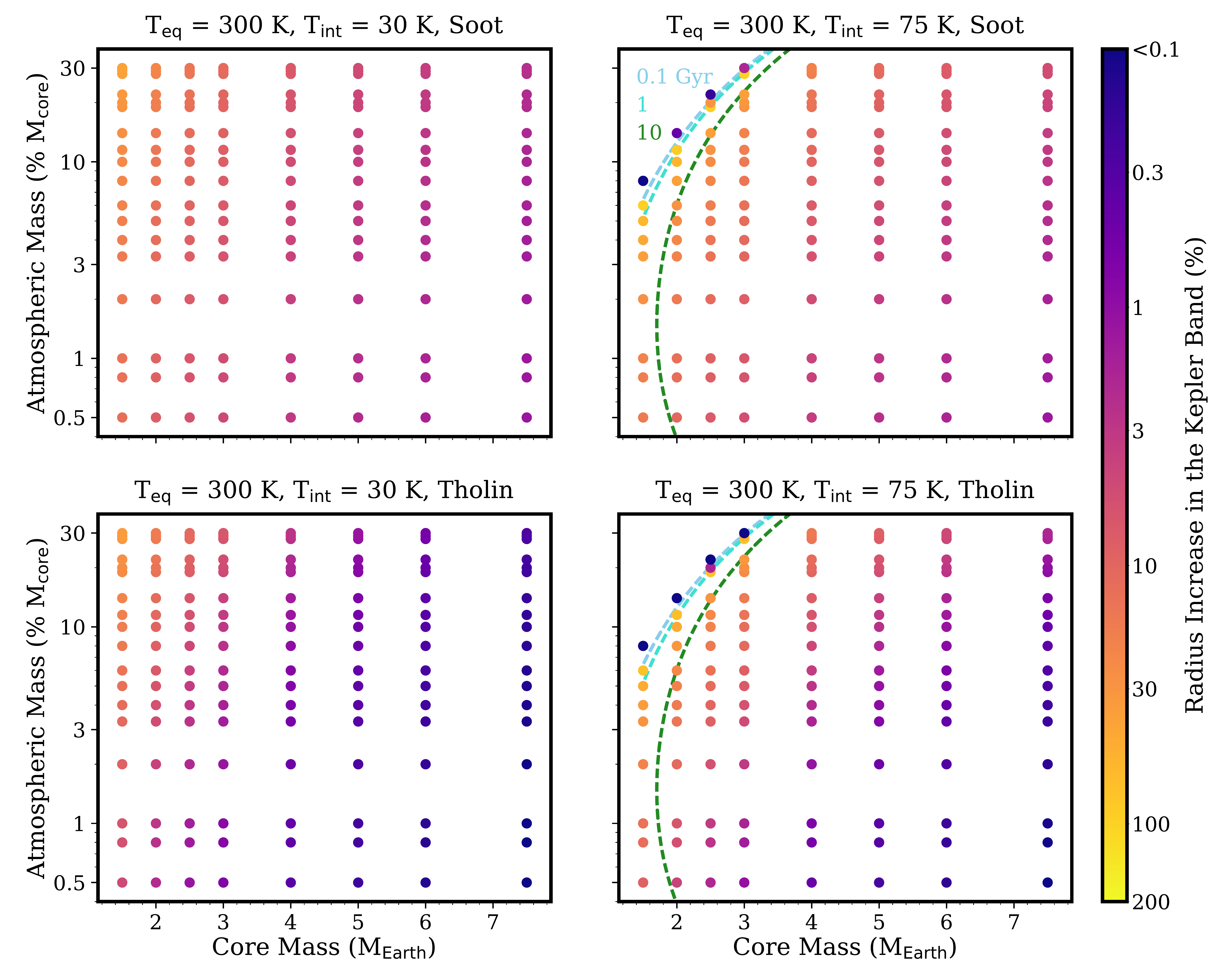}
\caption{Same as Figure \ref{fig:teq500raddif}, but for 300 K cases.}
\label{fig:teq300raddif}
\end{figure*}

\section{Tholin Haze Mass-Radius Diagram}\label{sec:appthomassrad}

\begin{figure*}[hbt!]
\centering
\includegraphics[width=0.7 \textwidth]{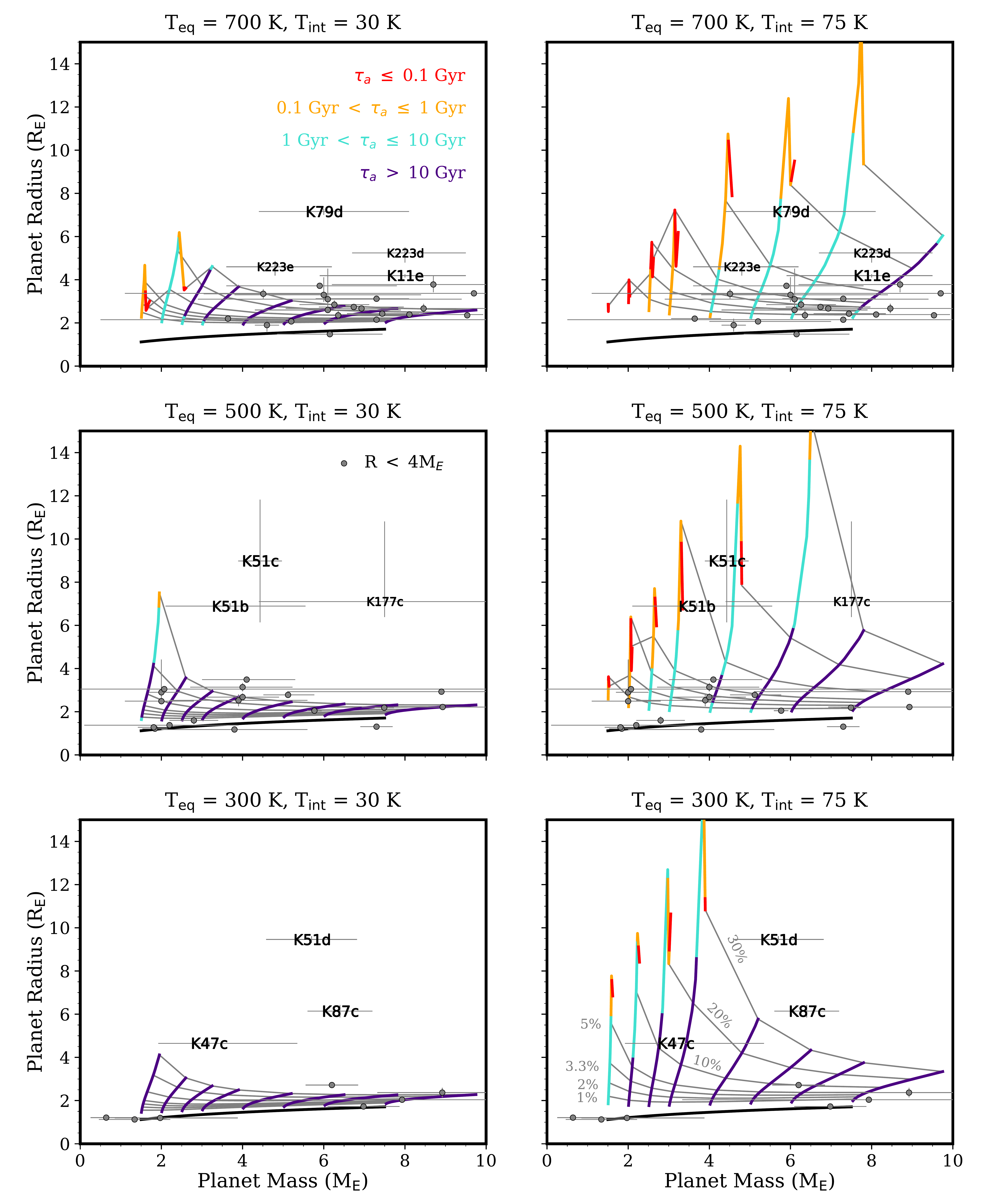}
\caption{Same as Figure \ref{fig:clearmr}, but for planets with tholin hazes. }
\label{fig:tholinmr}
\end{figure*}

\section{1.4 $\mu$m Water Feature Amplitude at 700 and 300 K}\label{sec:apph2oamp}

\begin{figure*}[hbt!]
\centering
\includegraphics[width=0.7 \textwidth]{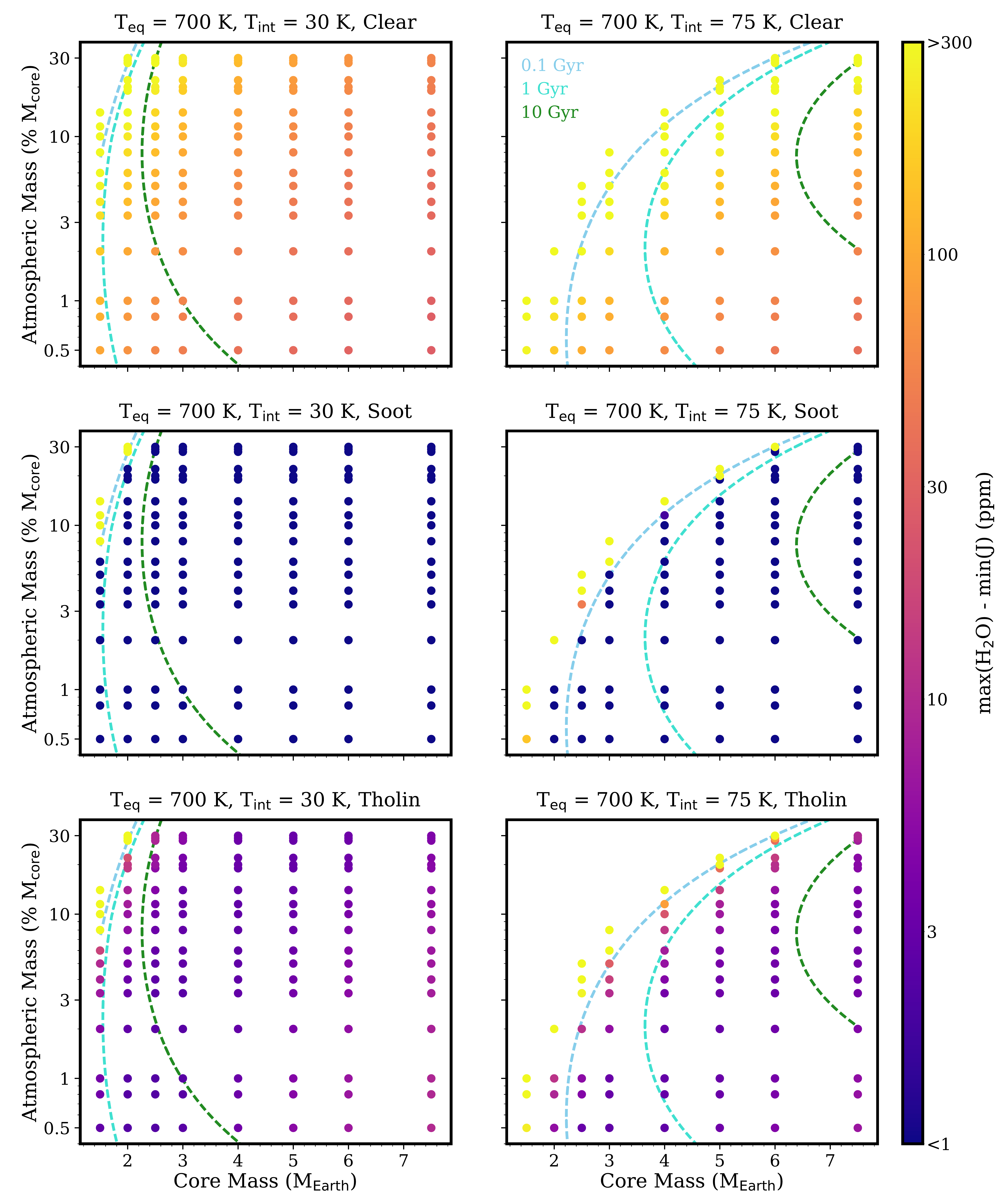}
\caption{Same as Figure \ref{fig:teq500h2oamp}, but for 700 K cases.}
\label{fig:teq700h2oamp}
\end{figure*}

\begin{figure*}[hbt!]
\centering
\includegraphics[width=0.7 \textwidth]{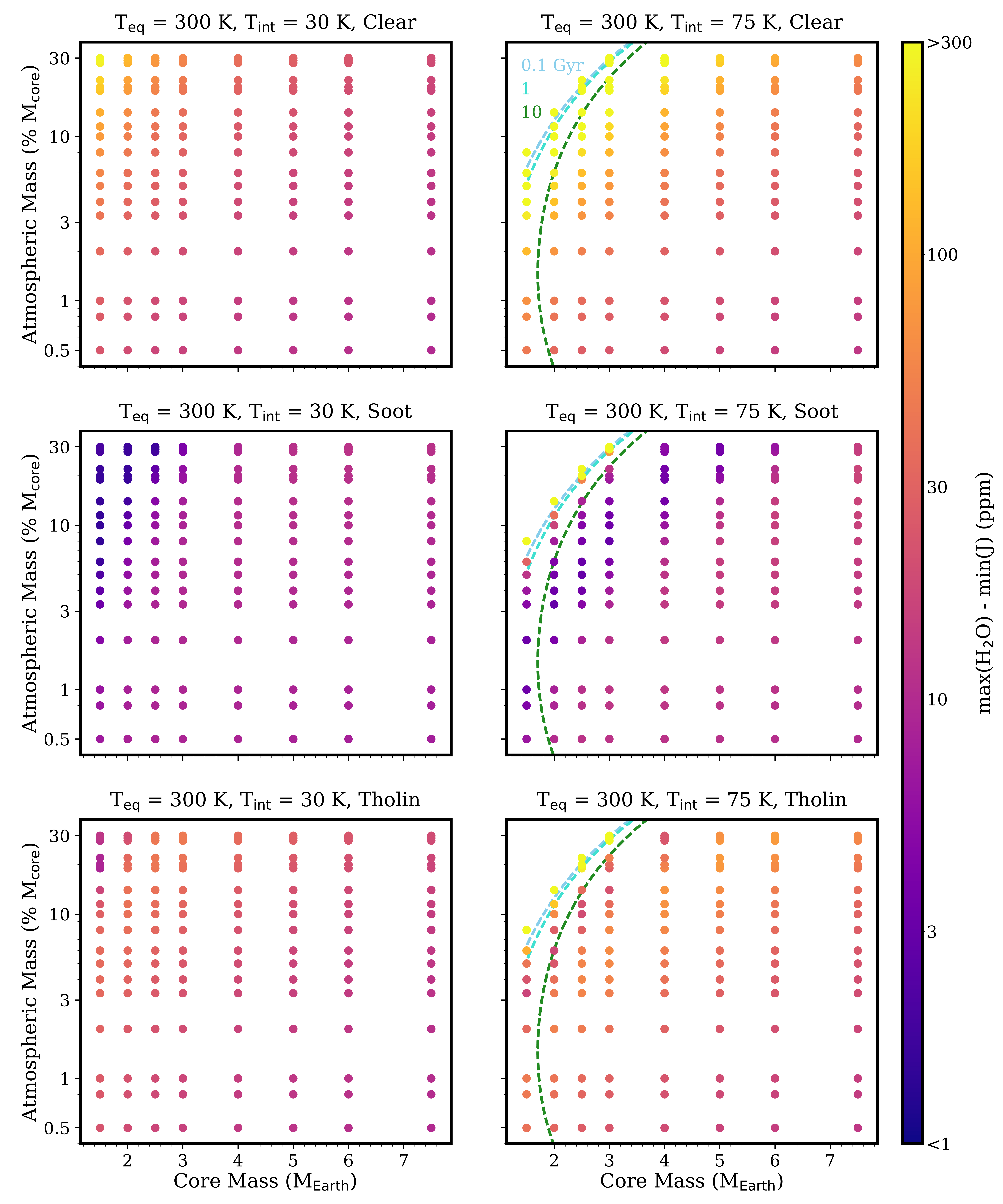}
\caption{Same as Figure \ref{fig:teq500h2oamp}, but for 300 K cases.}
\label{fig:teq300h2oamp}
\end{figure*}

\end{document}